\documentclass[longauth]{aa}  

\usepackage{graphicx}
\usepackage{txfonts}
\usepackage{multirow}
\usepackage{gensymb}
\usepackage{hyperref}
\hypersetup{breaklinks=true,colorlinks=true,urlcolor=blue, linkcolor=blue, citecolor=blue}

\begin{document}

   \title{A SPHERE survey of self-shadowed planet-forming disks \thanks{Based on observations collected at the European Southern Observatory under ESO programs: 096.C-0248, 097.C-0523, 0100.C-0452, 0101.C-0383, 0101.C-0464, 0102.C-0165, 198.C-0209, 1100.C-0481, and 1104.C-0415.}}

   \author{A.\,Garufi \inst{\ref{Arcetri}}
   \and C.\,Dominik \inst{\ref{Amsterdam}}
   \and C.\,Ginski \inst{\ref{Amsterdam}}
   \and M.\,Benisty \inst{\ref{CNRS}, \ref{IPAG}}
   \and R.G.\,van Holstein \inst{\ref{Leiden}}
   \and Th.\,Henning \inst{\ref{MPIA}}
   \and N.\,Pawellek \inst{\ref{Cambridge}}
   \and C.\,Pinte \inst{\ref{IPAG}}
   \and H.\,Avenhaus \inst{\ref{Klagenfurt}}
   \and S.\,Facchini \inst{\ref{UniMi}, \ref{ESO}}
   \and R.\,Galicher \inst{\ref{LESIA}}
   \and R.\,Gratton \inst{\ref{Padova}}
   \and F.\,M\'{e}nard \inst{\ref{IPAG}}
   \and G.\,Muro-Arena \inst{\ref{Amsterdam}}
   \and J.\,Milli \inst{\ref{IPAG}}
   \and T.\,Stolker \inst{\ref{ETH}}
   \and \\ A.\,Vigan \inst{\ref{LAM}}
   \and M.\,Villenave \inst{\ref{JPL}, \ref{IPAG}}
   \and T.\,Moulin \inst{\ref{IPAG}}
   \and A.\,Origne \inst{\ref{LAM}}
   \and F.\,Rigal \inst{\ref{Amsterdam}}
   \and J.-F.\,Sauvage \inst{\ref{ONERA}, \ref{LAM}}
   \and L.\,Weber \inst{\ref{Geneva}}
   }

  \institute{INAF, Osservatorio Astrofisico di Arcetri, Largo Enrico Fermi 5, I-50125 Firenze, Italy. \label{Arcetri}
  \email{antonio.garufi@inaf.it}  
  \and Anton Pannekoek Institute for Astronomy, University of Amsterdam, Science Park 904,1098XH Amsterdam, The Netherlands \label{Amsterdam}
  \and Unidad Mixta Internacional Franco-Chilena de Astronom\'{i}a (CNRS UMI 3386), Departamento de Astronom\'{i}a, Universidad de Chile, Camino El Observatorio 33, Las Condes, Santiago, Chile \label{CNRS}
  \and Univ. Grenoble Alpes, CNRS, IPAG, F-38000 Grenoble, France \label{IPAG}
  \and Leiden Observatory, Leiden University, PO Box 9513, 2300 RA Leiden, The Netherlands \label{Leiden}
  \and Max Planck Institute for Astronomy, K\"onigstuhl 17, D-69117 Heidelberg, Germany \label{MPIA} 
  \and Institute of Astronomy, University of Cambridge, Madingley Road, CB3 0HA, Cambridge, UK \label{Cambridge}
  \and Lakeside Labs, Lakeside Park B04b, 9020 Klagenfurt, Austria \label{Klagenfurt}
  \and Dipartimento di Fisica, Universit\`{a} degli Studi di Milano, via Celoria 16, 20133 Milano, Italy \label{UniMi}
  \and European Southern Observatory, Karl-Schwarzschild-Str. 2, 85748 Garching bei M\"unchen, Germany \label{ESO} 
  \and LESIA, Observatoire de Paris, Universit\'{e} PSL, CNRS, Sorbonne Universit\'{e}, Universit\'{e} de Paris, 5 place Jules Janssen, 92195 Meudon, France \label{LESIA}
  \and  INAF-Osservatorio Astronomico di Padova, Padova, Italy \label{Padova}
  \and Institute for Particle Physics and Astrophysics, ETH Zurich, Wolfgang-Pauli-Strasse 27, 8093 Zurich, Switzerland \label{ETH}
  \and Aix Marseille Univ, CNRS, CNES, LAM, Marseille, France \label{LAM}
  \and Jet Propulsion Laboratory, California Institute of Technology, 4800 Oak Grove Drive, Pasadena, CA 91109, USA \label{JPL}
  \and DOTA, ONERA, Universit\'{e} Paris Saclay, F-91123, Palaiseau France\label{ONERA}
  \and Geneva Observatory, University of Geneva, Chemin des Mailettes 51, 1290 Versoix, Switzerland \label{Geneva}
             }

   \date{Received -; accepted -}

% \abstract{}{}{}{}{} 
% 5 {} token are mandatory
 
  \abstract{To date, nearly two hundred planet-forming disks have been imaged at high resolution. Our propensity to study bright and extended objects does, however, bias our view of the disk demography. In this work, we aim to help alleviate this bias by analyzing fifteen disks targeted with VLT/SPHERE that look faint in scattered light. Sources were selected based on a low far-infrared excess from the spectral energy distribution. The comparison with the ALMA images {available for a few sources} shows that the scattered light {surveyed by these datasets} is only detected from a small portion of the disk extent. The mild anticorrelation between the disk brightness and the near-IR excess demonstrates that these disks are self-shadowed: the inner disk rim intercepts much starlight and leaves the outer disk in penumbra. Based on the uniform distribution of the disk brightness {in scattered light} across all spectral types, self-shadowing would act similarly for inner rims at a different distance from the star. We discuss how the illumination pattern of the outer disk may evolve with time. Some objects in the sample {are proposed to be} at an intermediate stage toward bright disks from the literature, with either no shadow or with signs of azimuthally confined shadows.}


\authorrunning{Garufi et al.}

\titlerunning{A SPHERE survey of self-shadowed planet-forming disks}

   \maketitle
%
%-------------------------------------------------------------------
\section{Introduction}
The high-contrast and high-resolution imaging enabled by the latest generation instruments has revolutionized our knowledge of planet-forming disks. As of today, resolved images of nearly 200 such disks are available thanks for example to the Hubble Space Telescope (HST), the Gemini Planet Imager (GPI), the Spectro-Polarimetric High-contrast Exoplanet REsearch (SPHERE) at the Very Large Telescope (VLT), and the Atacama Large Millimeter/submillimeter Array (ALMA). This census has allowed us to determine the diverse physical properties of planet-forming disks with a large range of different radial and vertical extents, temperatures and masses, as well as a variety of substructures \citep[cavities, rings, spirals, e.g.,][]{Muto2012, Benisty2015, Andrews2018}. It is, however, increasingly clear that the current census is severely biased toward massive, bright, and extended disks \citep[e.g.,][]{Garufi2018}. 

\begin{figure*}
  \centering
 \includegraphics[width=18cm]{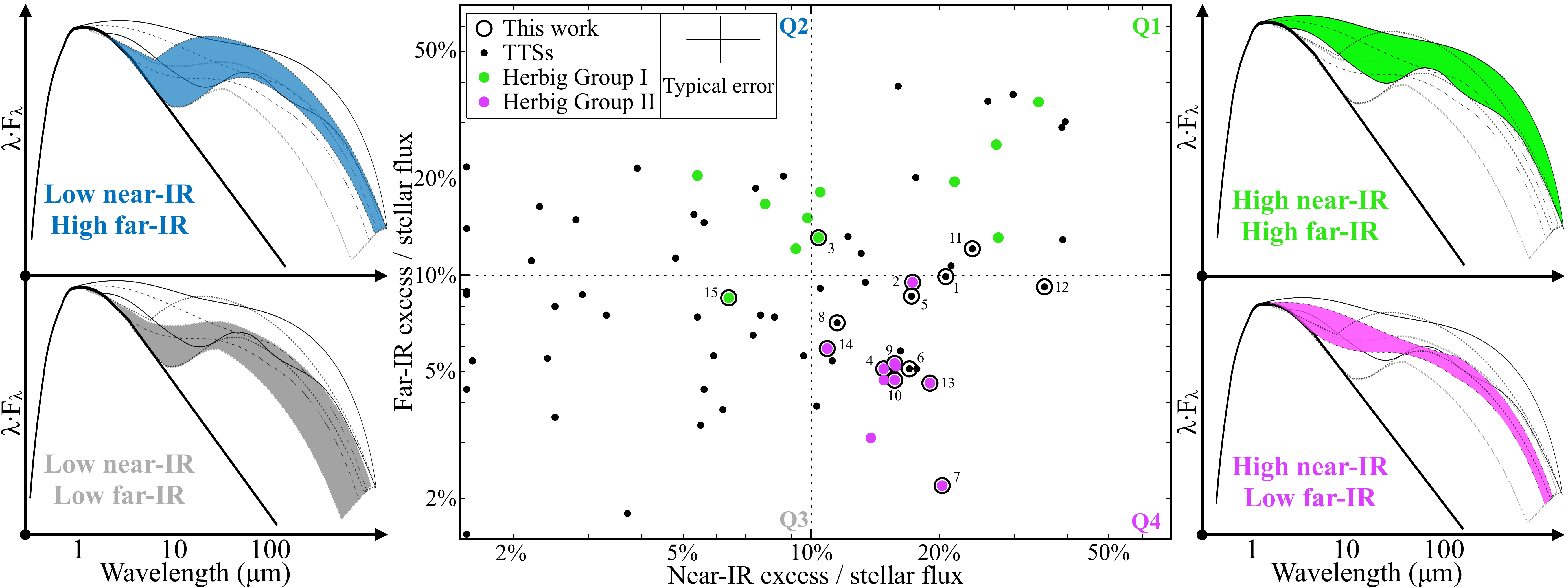} 
      \caption{NIR vs. FIR excess for an illustrative sample of stars with VLT/SPHERE images available. Both excesses are shown in fraction of the stellar luminosity. The dashed lines indicate the ideal separation between low and high excesses determining four quadrants. The range of IR excess exhibited by the targets of each quadrant is shown laterally. Target numbering refers to Table \ref{Stellar_properties}.}
          \label{FIR_NIR}
  \end{figure*}

The first sample of imaged disks to be assembled mainly consisted of Herbig Ae/Be stars {\citep[e.g.,][]{Grady1999, Fukagawa2006, Quanz2011}}. A classical divide between the spectral energy distribution (SED) of Herbig stars is the two groups first introduced by \citet{Meeus2001}. The Group II sources in this classification show an IR excess that can essentially be described by a single power law, while the Group I sources show an additional, very strong black-body component in the far-IR (FIR). The FIR component is commonly believed to originate from the flaring surface of the outer disk (>10 au), and its absence in Group II targets would suggest a poorly illuminated and possibly flatter disk surface \citep{Dullemond2004a}. {Instead}, the disk inner rim at a sub-astronomical-unit scale is directly exposed to the nearby stellar radiation. Therefore, being hot and puffed up, this disk region can cast an extended shadow on the outer disk \citep{Dullemond2001, Dong2015c} and leaves the characteristic near-IR (NIR) bump in the SED that is often observed in Herbig stars \citep[with amounts spanning from 5\% to 30\% of the stellar flux, e.g.,][]{Natta2001, Meeus2001}. In this context, an evolutionary link from flaring Group I to self-shadowed Group II disks because of dust settling in the outer regions has been put forward \citep[e.g.,][]{Dullemond2004b}.

More recently, it has been established from high-resolution observations that all Group I sources host a disk with a large cavity \citep{Maaskant2013, Garufi2014b, Honda2015}, where the inner rim of the outer disk -- typically located at 10--50 au -- is directly illuminated by the star, and thus it generates the observed FIR black-body component in the SED. Therefore, this cavity appears to either cause or at least to be connected with the disk flaring structure, which is why the Group I objects have been the preferred targets for all the initial NIR imaging campaigns and have routinely been detected by such studies. {Instead}, Group II sources have either no evidence for a cavity, or only interferometric evidence for very small cavities \citep{Menu2015}. Several studies have shown that Group II disks are faint in scattered light, as high-contrast images of such disks reveal little or no signal \citep[e.g.,][]{Grady2005, Garufi2017}. The physical explanations that have been proposed include either strongly settled and shadowed disks or the possibility that such disks are small compared to the angular resolution achieved by current telescopes. In fact, several ALMA surveys revealed that a large fraction of disks in star-forming regions are smaller than 30 au \citep[e.g.,][]{Barenfeld2017, Cox2017,Cieza2019}. In view of these findings, the evolution from Group I to Group II has been brought into question. Disks with increasingly large or deep inner cavities may turn from Group II to Group I, or, alternatively, disks with cavities formed early in their evolution \citep[see e.g.,][]{Sheehan2017} may represent a different evolutionary path from self-shadowed or small disks that are unable to create large cavities.

In this paper, we present NIR observations of several Herbig Group II sources or objects with an intermediate classification. We also extend the classification to {the later G-, K-, and M-type} T Tauri stars (TTSs), and we target those TTSs with the SED properties peculiar of the Herbig Group II. The VLT/SPHERE observations that we studied probe the linearly polarized scattered light from the planet-forming disk and are thus very sensitive to the disk illumination. Beside the routinely studied resolved images, polarimetric observations of young stars also allow us to infer the polarization properties of the unresolved flux in the stellar proximity \citep[see][]{Keppler2018, vanHolstein2020, Garufi2020a} and therefore to infer the presence and the basic properties of disks smaller than the angular resolution ($\lesssim$10 au).

The paper is organized as follows. In Sect.\,\ref{Nomenclature}, we extend the concept of Herbig groups to TTSs and present the sample of unpublished observations. In Sect.\,\ref{SPHERE}, we describe and examine the VLT/SPHERE observations, while in Sects.\,\ref{ALMA} and \ref{Correlations} we relate them to literature ALMA observations and other stellar and disk properties. Finally, in Sects.\,\ref{Discussion} and \ref{Conclusions} we discuss and summarize our findings.
   
%--------------------------------------------------------------------

\section{Nomenclature and sample} \label{Nomenclature}

\subsection{Group I and Group II for TTSs} \label{Extension_TTSs}
{Herbig Ae/Be stars and TTSs were initially identified by the presence of strong emission lines in their optical spectra and by their common association with obscured regions and bright nebulosities \citep{Herbig1960, Herbig1962}. Their divide is determined by the spectral type with TTSs being later F-, or G-, K-, and M-type stars. Low-mass stars ($\rm <1\,M_\odot$) exhibit the spectrum of TTSs right from their earliest stages, while intermediate-mass stars ($\rm 1-5\,M_\odot$) evolve from TTSs (starting from K-type) to Herbig Ae stars \citep[see e.g.,][]{Calvet2004}. Therefore, only focusing on either Herbig or TTSs determines not only a specific stellar mass interval for the observed sample, but also a specific age parameter space.}

The separation of Herbig stars into Group I (hereafter GI) and Group II (GII) is solely based on the properties of their SED \citep[see e.g.,][]{vanBoekel2003, Acke2006}. Here, we extend this classification to TTSs to broaden the connection between SED properties and disk geometry. However, we caution that the irradiation effect of these two types of stars is different and this may have an impact on the SED properties that is not grasped by the taxonomical divide in question.

\begin{table*}
      \caption[]{Stellar and disk properties of the targets of this work, {sorted by right ascension}.}
         \label{Stellar_properties}
     $$ 
         \begin{tabular}{lccccccccccc}
            \hline
            \hline
            \noalign{\smallskip}
            Source & Ref. & $d$ & T$_{\rm eff}$ & $A_{\rm V}$ & Group & $L_*$ & $M_*$ & $t$ & $M_{\rm d}$ & NIR & FIR \\
             & n. & (pc) & (K) & (mag) & & (L$_{\odot}$) & (M$_{\odot}$) & (Myr) & (M$_{\oplus}$) & (\%) & (\%)  \\
             \hline
             \noalign{\smallskip}
    CI Tau & 1 & 137.0 & 4400 & 1.9 & I/II & 1.3$\pm$0.2 & 1.1$\pm$0.2 & 1.5$-$5 & 135 & 20.7 & 9.9  \\
    HD287823 & 2 & 347.2 & 8375 & 0.0 & I/II & 11.7$\pm$0.7 & 1.8$\pm$0.2 & 8$-$16 & - & 17.3 & 9.5  \\
    HD245185 & 3 & 414.9 & 10000 & 0.4 & I & 26.7$\pm$1.3 & 2.1$\pm$0.2 & 7$-$11 & 50 & 10.4 & 13.1 \\
    HD290770  & 4 & 398.4 & 10500 & 0.0 & II & 34.3$\pm$1.9 & 2.2$\pm$0.2 & 6$-$15 & - & 14.8 & 5.1 \\
    TW Cha & 5 & 183.1 & 3955 & 0.8 & I/II & 0.64$\pm$0.04 & 1.1$\pm$0.1 & 1.5$-$5 & 24 & 17.2 & 8.6  \\
    DI Cha & 6 & 189.0 & 5860 & 1.8 & II & 10.5$\pm$0.8 & 2.3$\pm$0.2 & 0.5$-$2 & 10 & 17.0 & 5.1 \\
    HD98922  & 7 & 653.6 & 10500 & 0.1 & II & 1219$\pm$63 & 5.5$\pm$0.6 & 0.2$-$0.7 & - & 20.3 & 2.2  \\
    GW Lup & 8 & 155.3 & 3630 & 0.8 & II & 0.43$\pm$0.06 & 0.7$\pm$0.1 & 1$-$3 & 50 & 11.5 & 7.1  \\
    HD142666 & 9 & 146.4 & 7500 & 0.9 & II & 11.9$\pm$1.3 & 1.7$\pm$0.2 & 7.5$-$11.5 & 44 & 15.7 & 5.3  \\
    HD144432 & 10 & 155.0 & 7500 & 0.4 & II & 13.7$\pm$0.9 & 1.7$\pm$0.2 & 6.5$-$10.5 & 30 & 15.7 & 4.7  \\
    V1003 Oph & 11 & 114.7 & 5770 & 4.5 & I/II & 7.1$\pm$0.8 & 1.9$\pm$0.2 & 2$-$6 & 10 & 23.9 & 12.1  \\
    WSB82 & 12 & 146.0 & 4800 & 4.5 & I/II & 1.5$\pm$0.4 & 1.4$\pm$0.2 & 3$-$7 & 120 & 35.3 & 9.2  \\
    HD150193 & 13 & 150.8 & 9000 & 1.5 & II & 26.7$\pm$4.8 & 2.2$\pm$0.2 & 4$-$7 & 24 & 19.0 & 4.6  \\
    AK Sco & 14 & 139.9 & 6250 & 0.7 & II & 2 $\times$ (3.2$\pm$0.6) & 2 $\times$ (1.3$\pm$0.2) & >12 & 16 & 10.9 & 5.9  \\
    HD179218  & 15 & 260.4 & 9640 & 0.4 & I/II & 115$\pm$8 & 2.9$\pm$0.2 & 1.5$-$3.5 & 140 & 6.4 & 8.5  \\
    \hline
    \end{tabular}
     $$ 
     \tablefoot{Columns are: source name, reference number in figures, distance, effective temperature, optical extinction, group based on Fig.\,\ref{FIR_NIR}, stellar luminosity, mass, age, dust mass in the disk, {NIR and FIR excess normalized to the stellar luminosity}. Distances are from Gaia EDR3 \citep{Gaia2021}. Effective temperatures of Herbig stars are from \citet{Fairlamb2015}, of CI Tau from \citet{Herczeg2014}, of GW Lup from \citet{Alcala2017}, of TW Cha and DI Cha from \citet{Luhman2007}, and of V1003 Oph and WSB82 from \citet{Luhman1999}. The other properties are calculated in this work as described in Appendix \ref{appendix_properties}. The dust mass in the disk is obtained under standard assumptions (see Appendix \ref{appendix_properties}) and should be regarded as an indicative value.}
   \end{table*}

The most recent criterion to be proposed to define Herbig GI and GII involves the slope of the mid-IR SED. The [30 $\mu$m/13.5 $\mu$m] flux ratio (herafter [30/13]) can be used as a direct measurement of the warm dust continuum \citep{Acke2009} that leaves the partition of the two groups substantially unaltered \citep{Maaskant2013}. %Following this criterion, \citet{Garufi2017} adopted the threshold [30/13]=2.2 since this value corresponds to a flat SED.
In fact, the Herbig stars with [30/13] larger or smaller than 2.2 (GI or GII) exhibit a FIR excess larger or smaller than 10\% of the stellar luminosity \citep{Garufi2017} and thus reflect the original classification by \citet{Meeus2001}. However, this criterion cannot be extended to TTSs since the relation between the mid-IR color and the FIR excess is no longer present. Several TTSs are, in fact, partly embedded in the natal envelope or lie in the dense environment of star-forming regions. The SEDs of these sources show a globally high IR excess resulting in a rather small [30/13] color despite the large FIR excess.

Alternatively, Herbig GII sources can also be defined by the two-dimensional parameter space of NIR and FIR excess. All Herbig stars with [30/13]<2.2 (GII) show a NIR excess within a narrow interval of values \citep[14\%$-$19\%, see][]{Banzatti2018}. {Instead}, sources with [30/13]>2.2 (GI) show a bimodal distribution with either low NIR ($\lesssim$10\%) or high NIR ($>$20\%). Therefore, setting a value of 10\% to separate high and low excesses in a NIR or FIR diagram (Fig.\,\ref{FIR_NIR}) yields the divide of Herbig stars in three clusters, namely high-NIR GI, low-NIR GI, and GII \citep[as in][]{Banzatti2018}. 

Based on these premises, we took an illustrative sample of 55 TTSs with VLT/SPHERE polarimetric images available and calculated their NIR and FIR excess {from 1.2 to 4.5 $\mu$m and from 22 to 450 $\mu$m, respectively,} in analogy with \citet[see also Appendix \ref{appendix_properties}]{Garufi2017}. Figure \ref{FIR_NIR} shows that TTSs are more uniformly distributed in the diagram than Herbig stars. Many TTSs are found with low NIR excess and moderate FIR excess (quadrant {Q3}), where Herbig stars are uncommon. In any case, those TTSs sitting close to the three Herbig clusters can formally be defined as the TTS Group I and Group II counterpart of Herbig stars. This diagram is further discussed in Sect.\,\ref{Discussion_evolution} in view of the results of this work.

\subsection{Sample}
The main sample of this work consists of {nine} Herbig and {six} TTSs with unpublished VLT/SPHERE observations from different programs (see Appendix \ref{appendix_observations}) that are classified, according to the conclusions of Sect.\,\ref{Extension_TTSs}, as GII or intermediate between GI and GII. As is clear from Fig.\,\ref{FIR_NIR}, {six} of these clearly fall in the quadrant {Q4} (that of GII). These are HD142666, HD144432, HD150193, HD290770, HD98922, and DI Cha. The first five of these are prototypical Herbig GII stars, while, based on the arguments of Sect.\,\ref{Extension_TTSs}, DI Cha is a TTS GII source. {Two sources, namely HD152404 (AK Sco) and GW Lup (Sz71), lie between Q3 and Q4.} AK Sco is an intermediate object that shows the [30/13] color of GI and the FIR of GII \citep[see Fig.\,3 of][]{Garufi2017}. {Conversely}, five targets of the sample fall in the transition region between high-NIR GI and GII ({Q1 and Q4}). These are CI Tau, HD287823, TW Cha, V1003 Oph, and WSB82. Finally, HD179218 and HD245185 could be considered intermediate cases, similarly to AK Sco but with an opposite trend, showing the [30/13] color of GII and the integrated excesses of GI.   

The stellar properties of all targets are listed in Table \ref{Stellar_properties}. The stellar luminosity and dust mass in the disk are updated {after Gaia EDR3 \citep{Gaia2021} from} \citet{Garufi2018}, or they are calculated analogously when the source is not in the original sample. Stellar masses and ages are obtained from stellar isochrones as described in Appendix \ref{appendix_properties}. From Table \ref{Stellar_properties}, it is clear that the sample mainly consists of intermediate-mass stars in the 1$-$3 M$_\odot$ mass range, with the exception of GW Lup (<1 M$_\odot$) and HD98922 ($\sim$5 M$_\odot$). The calculated ages span from around 1 Myr (HD98922 and DI Cha) to more than 10 Myr (HD142666 and AK Sco). Dust masses in the disk span more than one order of magnitude, from 10 M$_\oplus$ to 140 M$_\oplus$. Five sources are located at more than 250 pc, while the other ten sources are closer than 190 pc. 

\begin{figure*}
  \centering
 \includegraphics[width=18cm]{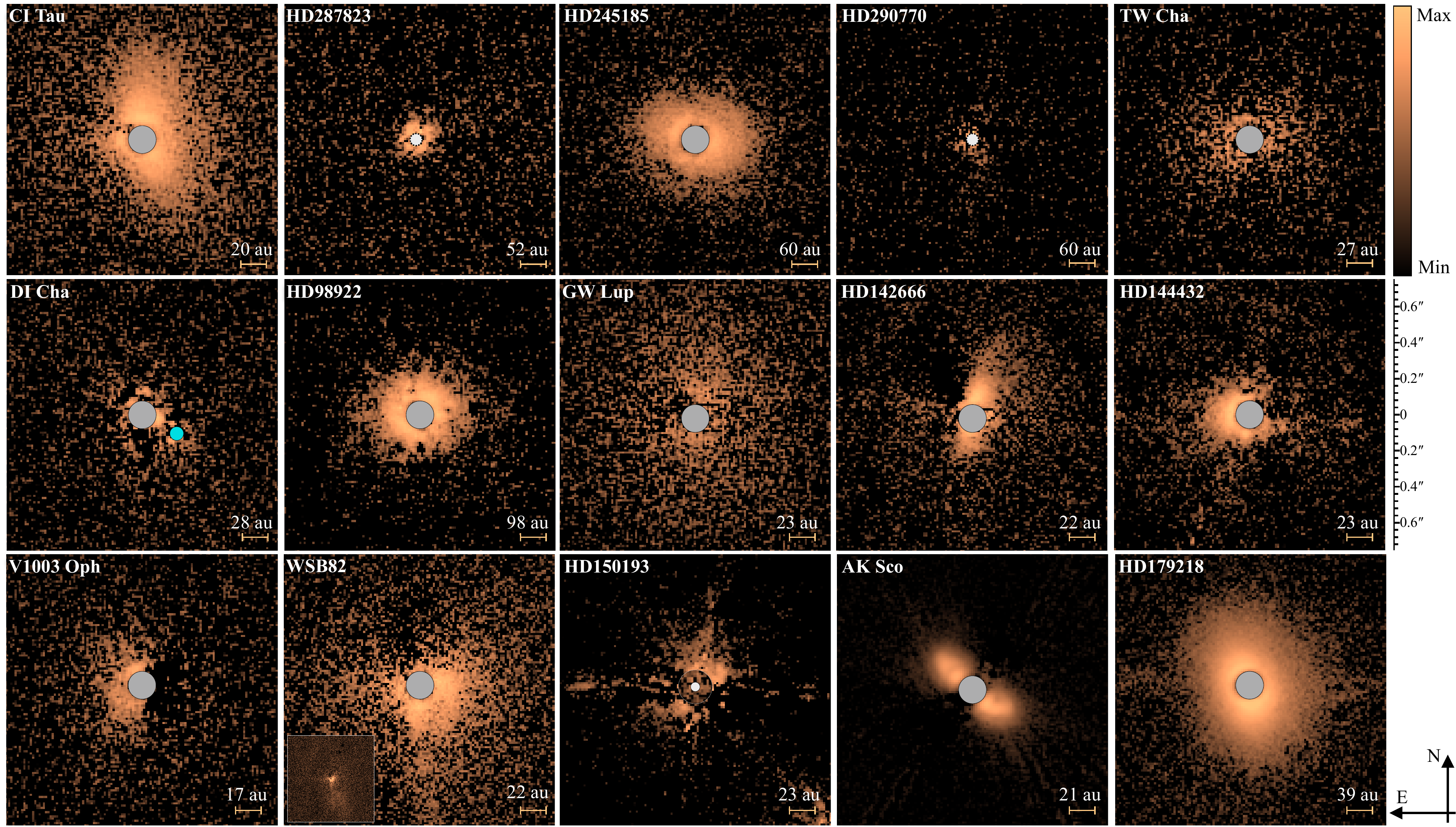} 
      \caption{Imagery of the sample. For each source, the $Q_{\phi}$ map is shown with an arbitrary logarithmic color stretch. {The relative brightness of the individual disks can be evaluated from Fig.\,\ref{Contrast_list}. All images have the same angular field (1.5\arcsec), and the physical scale is indicated by the bar in each panel.} The gray circle in the center of a map indicates the coronagraph. Smaller sized white circles denote the angular resolution of non-coronagraphic images. Stellar companions in the field are indicated by cyan circles. The image of HD150193 is a combination of coronagraphic {J-band (main image)} and non-coronagraphic {H-band (inner inset at 0.2\arcsec)} images. {The diffraction spikes in this source are evident toward the four cardinal points. The inset image of WSB82 shows a larger field (7\arcsec) of the same image.} }
          \label{Imagery}
  \end{figure*}

\section{VLT/SPHERE images} \label{SPHERE}

\subsection{Observing mode and data reduction} \label{Observations}
All the observations studied in this work were obtained in the NIR with the Infra-Red Dual Imaging and Spectrograph \citep[IRDIS,][]{Dohlen2008} of SPHERE \citep{Beuzit2019} in dual-beam polarimetric imaging mode \citep[DPI,][]{Langlois2014, vanHolstein2020, deBoer2020}. This mode allows us to separate the dominant NIR stellar light and the polarized light from the stellar surrounding producing high-resolution and high-contrast maps of the light scattered by the disk. The observations were carried out in different programs and with slightly different setups (see Appendix \ref{appendix_observations} for details). Four targets were observed in the broad $J$ band {(at 1.245 $\mu$m with a width of 0.24 $\mu$m)} of SPHERE, two in the narrow $H$ band {(at 1.573 $\mu$m with a width of 0.02 $\mu$m)}, and the remaining nine in the broad $H$ band {(at 1.625 $\mu$m with a width of 0.29 $\mu$m)}. In one case, HD150193, both broad $J$ and narrow $H$ bands are available. All observations taken in broad band were aided by an apodized Lyot coronagraph (ALC) of either 145 mas (N\_ALC\_YJ\_S) or 185 mas (N\_ALC\_YJH\_S) in diameter. For the observations in the narrow band, no coronagraph was employed. Total integration times span from 13 to 85 minutes.

The total intensity and polarization state of the incoming light can be described by the Stokes formalism \citep[e.g.,][]{Tinbergen2005}. Positive and negative $Q$ Stokes parameters, $Q^+$ and $Q^-$, describe the vertical and horizontal linear polarization, while positive and negative $U$ Stokes parameters, $U^+$ and $U^-$, trace the component rotated by 45$\degree$ with respect to $Q^+$ and $Q^-$. From these parameters, the linearly polarized intensity is calculated as $PI=\sqrt{Q^2+U^2}$ , while the degree of linear polarization is  ${\rm DoLP}=\sqrt{q^2+u^2}$, with $q=Q/I$, $u=U/I$, and $I$ being the total intensity {obtained from the $I_Q$ and $I_U$ intensity images associated with the $Q$ and $U$ images.} Also, the angle of linear polarization is measured from ${\rm AoLP}=1/2\arctan{(u/q)}$. Most recent works on circumstellar disks make use of the azimuthal Stokes parameters \citep{deBoer2020} defined as $Q_\phi=-Q\cos{(2\phi)}-U\sin{(2\phi)}$ and $U_\phi=+Q\sin{(2\phi)}-U\cos{(2\phi)}$, where $\phi$ is the azimuth angle. In centro-symmetric scattering patterns such as those expected from moderately inclined disks, $Q_\phi$ corresponds to $PI$ with the benefit of not squaring the noise. However, part of $PI$ signal is expected to be transferred in the $U_\phi$ in disks with high inclination or in case of multiple scattering \citep{Bastien1988, Canovas2015}.

Data were reduced using the IRDAP\footnote{\url{irdap.readthedocs.io}} (IRDIS Data reduction for Accurate Polarimetry) pipeline, version 1.3.1 \citep{vanHolstein2020}. IRDAP contains a detailed Mueller matrix model of the SPHERE optical system that allows us to correct the stellar beam at the detector for the instrumental polarization without necessarily correcting for the intrinsic, unresolved polarization. Thus, the latter component can be included in the data analysis and investigated along with the final maps corrected for this type of polarization.

\subsection{Individual maps} \label{Individual}
The SPHERE $Q_\phi$ maps of the whole sample are shown in Fig.\,\ref{Imagery}. From a first look, it is clear that the signal in all maps appears faint, as is shown in Sect.\,\ref{Disk_brightness}. Here, we give a brief description of each object based on these maps. {The noise of each map is estimated from the same $Q_\phi$ images through the $\sigma$ of a resolution-large region devoid of any signal.}

The polarimetric image of CI Tau shows relatively bright signal in an {elliptical} shape that {is above 3$\sigma$} out to 0.75\arcsec\ ($\sim$100 au) north and south. {A fainter halo with a signal below $3\sigma$ can be retrieved out to $\sim$1.8\arcsec\ after smoothing the original image.} The signal is maximized along the 353\degree\ and 208\degree\ azimuthal angles. A comparison with ALMA images is given in Sect.\,\ref{ALMA}.

The polarimetric non-coronagraphic image of HD287823 only shows some very compact emission, with the outermost detectable signal at 0.09\arcsec\ (30 au) from the star. This extent is nearly three times larger than the resolution element in radius indicating a resolved emission. A hint of quadrupole pattern is visible from the Q and U images (see Appendix \ref{Appendix_QU}) suggesting that the signal in the $Q_\phi$ map is actually scattered light from circumstellar material on a small separation. 

Relatively bright polarimetric signal is {detected above 3$\sigma$} out to 0.5\arcsec\ (200 au) around HD245185. The flux is distributed in an ellipsoidal shape with a P.A.\ of 68\degree. %Along the minor axis, a flux discontinuity resembling the dark lane seen in brighter disks is visible (see XXX). 

The polarimetric non-coronagraphic map of HD290770 does not reveal any clear resolved signal. Unlike HD287823, the Q and U images do not exhibit any quadrupolar pattern (see Appendix \ref{Appendix_QU}). Nonetheless, the presence of unresolved signal detectable on a small scale is possible given the absence of any coronagraph (see Sect.\,\ref{Unresolved_pol}).

The polarimetric image of TW Cha shows very weak signal out to 0.3\arcsec\ (55 au). The signal is visible in the $Q$ and $U$ images (see Appendix \ref{Appendix_QU}), corroborating the detection of a relatively extended, faint disk. 

DI Cha is a quadruple system with a close companion (D) and a binary system (BC) orbiting at a greater distance \citep[see][and Sect.\,\ref{Stellar_companions}]{Schmidt2013}. The polarimetric image of Fig.\,\ref{Imagery} shows some signal to the SW in the vicinity of the coronagraph, but the Q and U images (see Appendix \ref{Appendix_QU}) do not reveal any compelling evidence of polarization.

The vaguely ellipsoidal shape of the polarized flux detected around HD98922 suggests a moderately inclined disk with a P.A.\ of approximately 100\degree. The flux outer extent along the putative disk major axis is 0.3\arcsec, translating, in view of the large distance of the source (650 pc) into a very large physical extent of {at least} 200 au in radius.

Weak polarized signal is detected around GW Lup mostly along the north-south direction. A comparison with ALMA images is given in Sect.\,\ref{ALMA}.

Polarized flux around HD142666 is clearly detected in a shape of two bright wings visible to the north and south, which suggests an inclined disk with a P.A.\ of approximately 170\degree. This disk geometry is consistent with the NACO image by \citet{Garufi2017} and the ALMA image by \citet{Andrews2018} (see Sect.\,\ref{ALMA}). 

The triple system HD144432 is composed of A-type (star A), K-type (B), and M-type (C) stars, where B and C closely orbit each other \citep[see][and Sect.\,\ref{Stellar_companions}]{Mueller2011}.  The polarimetric image of Fig.\,\ref{Imagery} reveals some signal to the east, in the immediate surrounding of the coronagraph. The Q and U images of this source (see Appendix \ref{appendix_observations}) show the quadrupole pattern expected for centro-symmetric scattering, indicating a real disk signal detected out to 0.25\arcsec\ ($\sim$40 au). The strong asymmetry between eastern and western sides may point to a half a disk being in shadow, like the protypical examples of HD143006 and HD139614 \citep{Benisty2018, MuroArena2020}. The interferometric data by \citet{Lazareff2017} and \citet{Perraut2019} constrained an inner disk P.A.\ of 76\degree$-$79\degree\ and a low inclination of 25\degree$-$30\degree.

The polarimetric map of V1003 Oph reveals positive signal in the eastern half around the coronagraph and negative signal in the western side. The average negative signal is half in the absolute value than the positive signal. Similarly to HD144432, such a morphology resembles that of partly shadowed disks (see Sect.\,\ref{Discussion}).

Relatively bright polarimetric signal is detected in WSB82 in the western and southern directions. This inner component is detectable out to $\sim$0.2\arcsec\ (30 au). Some diffuse, fainter signal is also detected at much larger separations (as far out as 3\arcsec, {see inset image in Fig.\,\ref{Imagery}}). A comparison with ALMA images is given in Sect.\,\ref{ALMA}.

Doubtful polarized signal may be detected from the coronagraphic polarimetric image of HD150193 in the NW and SE direction. In fact, the Q and U images (see Appendix \ref{Appendix_QU}) from both the coronagraphic and non-coronagraphic images show a hint of quadrupolar pattern. If the putative signal were real, the polarimetric image would portray scattered light from a relatively inclined disk with a P.A.\ of approximately 135\degree. The P.A.\ of the inner disk constrained from interferometric data spans from 131\degree\ to 176\degree\ \citep{Lazareff2017, Perraut2019, Kluska2020}, thus making it coarsely consistent. Nonetheless, this source is hereafter treated as a non-detection.

The disk of the AK Sco binary stars is clearly detected in polarized scattered light and is also visible in the total intensity image. Similarly to the polarization images by \citet{Garufi2017} and \citet{Esposito2020} as well as the intensity image by \citet{Janson2016}, the very inclined disk is seen with a P.A.\ of approximately 45$\degree$. The size of the bright visible disk is 0.25\arcsec\ while some signal is recovered out to 0.35\arcsec. The gap described by \citet{Esposito2020} on the NE disk side is also visible from our image and resembles a local shadow cast by the inner disk, as routinely imaged in scattered light \citep[e.g.,][]{Stolker2016a, Pinilla2018b}.

The disk of HD179218 was easily detected in scattered light. The spatial distribution of the detected signal is suggestive of a relatively inclined disk with a P.A.\ of approximately 15\degree. The signal is detected {above 3$\sigma$} out to 0.6\arcsec, translating into a rather large disk of at least 160 au in {radius}.

\subsection{Disk brightness in scattered light} \label{Disk_brightness}
The disk brightness in scattered light can be evaluated through the polarized-to-stellar light contrast $\delta_{\rm pol}$. With this, the natural dependence of the observed flux on the amount of stellar flux is canceled. The total amount of scattered light also depends on the scattering phase function \citep[see e.g.,][]{Mulders2013a}, on the disk extent, and on the projected area of the visible disk or, in other words, the disk inclination. These dependencies are significantly alleviated by computing the average value between the inner and outer radius with signal, on the disk major axis only, after normalizing at each separation by the squared distance \citep{Garufi2017}. 

{For this work,} the polarized flux per unit area $F_{\rm pol}$ {was extracted} along the disk major axis {in all the targets of Fig.\,\ref{Imagery}}. {The position angle of this axis may not be known for such faint disks, but since disks appear azimuthally symmetric in scattered light, this choice is of minor impact. Then, $F_{\rm pol}$} was multiplied by the squared separation $r$ (ignoring the minor impact of the disk scale height on the effective $r$) and averaged radially from the innermost disk radius $r_{\rm in}$ to the outermost radius with detectable signal $r_{\rm out}$. {Any signal at a separation smaller than the coronagraph size or angular resolution cannot be measured, and $r_{\rm in}$ is thus determined by the observing setup when the signal is traced down to the coronagraph or resolution element.} The value thus obtained is normalized by the stellar flux $F_*$ calculated from the \textsc{flux} frames taken during the SPHERE observations taking into account the different neutral density filters and exposure time of science and \textsc{flux} frames. In summary, the polarized-to-stellar light contrast can be expressed as:
\begin{equation} \label{Formula_albedo}
\delta_{\rm pol}= \frac{1}{r_{\rm out}-r_{\rm in}} \cdot \int_{r_{\rm in}}^{r_{\rm out}} \frac{F_{\rm pol}(r)\cdot 4\pi r^2}{F_*} dr 
.\end{equation}
The resulting measurement can be considered as a geometrical albedo expressing the fraction of stellar photons that are effectively scattered off (and polarized) {by the resolved portion of the disk (typically from a separation of 10--15 au, see also Sect.\,\ref{Unresolved_pol}). Formally speaking, any measured contrast is therefore a lower limit imposed by the finite resolution or by the employment of a coronagraph. While for large disks the missing portion of measured scattered light is small, in the case of small disks this can be very large up to the case of a disk smaller than $\sim$10 au, where the measured contrast of the resolved polarized-to-stellar light contrast will be zero.}

\begin{figure}
  \centering
 \includegraphics[width=9cm]{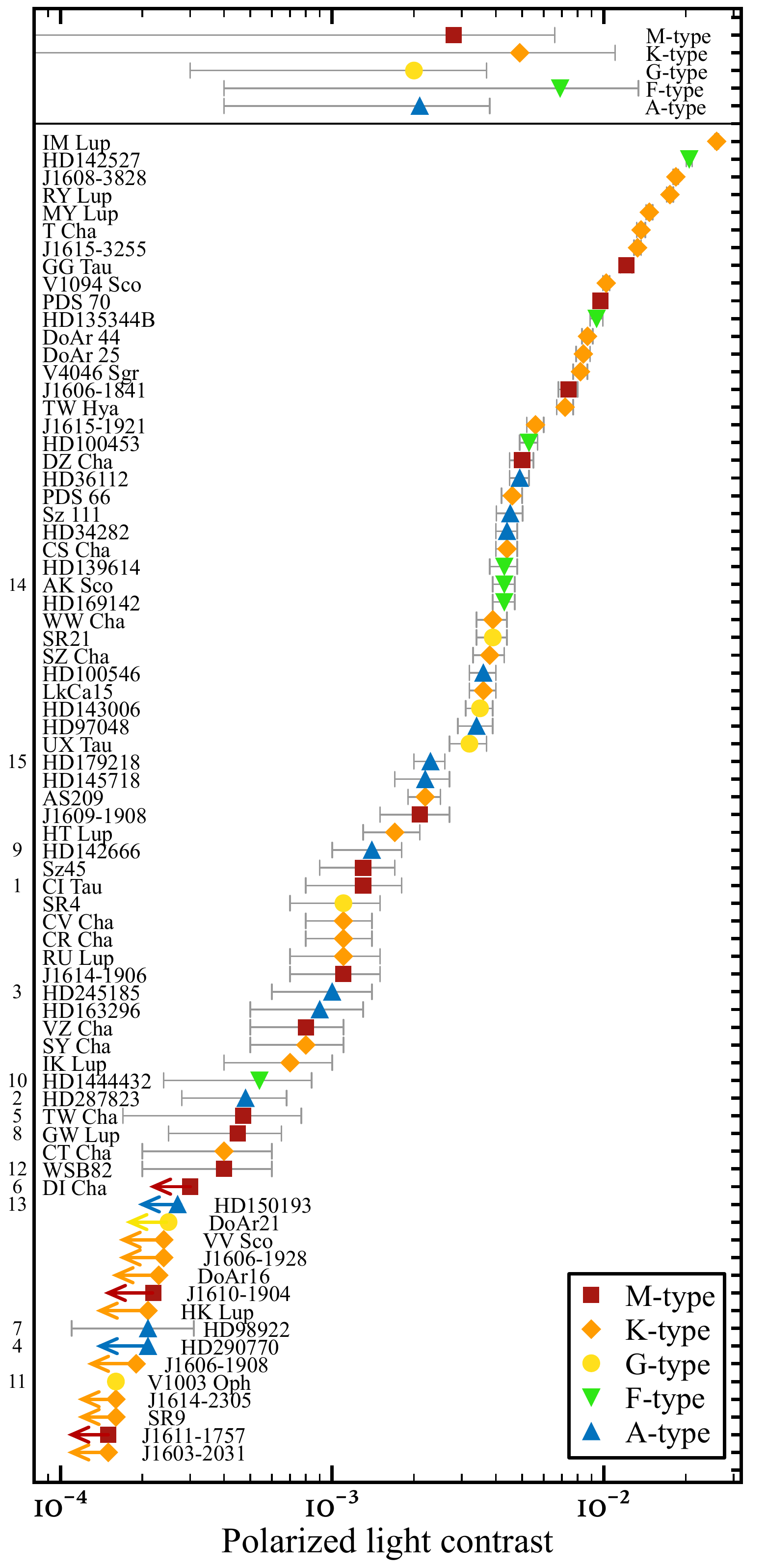} 
      \caption{Disk brightness in scattered light. The {resolved} polarized-to-stellar light contrast of the sample from this work is compared with an illustrative sample from the literature. Different symbols indicate different spectral types. For simplicity, B9 stars are combined with A stars. {The average of the spectral types is shown to the top.} The {numbers} to the left indicate the sources from this work {as from Table \ref{Stellar_properties}}. {Bars and arrows on the individual sources indicate 3$\sigma$ uncertainties and upper limits, respectively. The bars on the average of spectral types indicate the 1$\sigma$ dispersion.}}
          \label{Contrast_list}
  \end{figure}

The calculation of $\delta_{\rm pol}$ enables the direct comparison of different sources. In Fig.\,\ref{Contrast_list}, the $\delta_{\rm pol}$ calculated for the fifteen sources of Fig.\,\ref{Imagery} is compared with the benchmark targets of Fig.\,\ref{FIR_NIR}. The diagram reveals that the sample of this work is undoubtedly in the lower half of the distribution. In particular, the faintest disks (such as HD98922, WSB82, and GW Lup) are nearly two orders of magnitude fainter than the brightest disks (IM Lup, HD142527). On the other hand, the brightest disks from Fig.\,\ref{Imagery} (HD179218, AK Sco, CI Tau) are only 2$-$5 times fainter than the upper end of the distribution. Also, Fig.\,\ref{Contrast_list} does not reveal any obvious dependency for the contrast on the spectral type, {as is clear from the average values shown on top of the diagram}. It does, however, show that the very inclined disks (J1608-3828, RY Lup, MY Lup, T Cha) are all very bright. This is a bias due to the forward peak of the scattering phase function being probed in these objects and by the direct starlight from the photosphere being partly extincted by the disk. Therefore, the inclined disks of this work (HD142666 and AK Sco, see Fig.\,\ref{Imagery}) can legitimately be considered faint disks despite exhibiting the median $\delta_{\rm pol}$ of the distribution.

\subsection{Unresolved polarization} \label{Unresolved_pol}
Circumstellar material on separations smaller than the spatial resolution can still induce a measurable unresolved polarization on the observed stellar beam. For observations in the H band with an 8-m telescope, the ideal resolution is of the order of 50 mas (corresponding to, e.g., 7 au at 140 pc and 20 au at 400 pc). However, the stellar beam can also be polarized by interstellar dust and instrumental optics. While the latter component is corrected by IRDAP (see Sect.\,\ref{Observations}), the two components induced by circumstellar and interstellar material are entangled in the observed flux. Based on a sample of 21 targets, \citet{Garufi2020a} found that the angle of unresolved polarization traces the circumstellar disk orientation only in sources with low interstellar extinction ($A_{\rm V}\lesssim1.0$ mag), while for large interstellar extinction ($A_{\rm V}\gtrsim3.0$ mag) the interstellar material is the dominant source of polarization.

The degree and angle of unresolved polarization (DoLP and AoLP, see Sect.\,\ref{Observations}) calculated by IRDAP are shown in Table \ref{Pol_properties}. The five brightest disks in the sample (see Fig.\,\ref{Contrast_list}) all have a relatively large DoLP (>0.5\%). Their AoLP is perpendicular to the disk P.A., in line with the expectations \citep[see e.g.,][]{vanHolstein2020}. As for the two sources with asymmetric flux distribution (HD144432 and V1003 Oph, see Fig.\,\ref{Imagery}), the AoLP is perpendicular to the direction where the polarized flux is maximum. {Instead}, the high DoLP of WSB82, DI Cha, and HD150193 is possibly due to the interstellar material, given their large $A_{\rm V}$ (see Table \ref{Stellar_properties}).

The cases of HD287823 and HD290770 are particularly instructive. The disk around the former source is resolved (see Sect.\,\ref{Individual}), although the degree of unresolved polarization is very low (0.1\% and consistent with null). This indicates a low amount of polarizing material inside the region ($\sim$20 au in radius) where the disk is resolved, or that this material is uniformly distributed and on a small inclination inducing an azimuthally symmetric polarization that is canceled since unresolved. {Conversely}, the high DoLP (1\%) measured for a target, HD290770, with null interstellar extinction and no resolved signal indicates the substantial presence of polarizing material on scales smaller than 25 au only. These results are further discussed in Sect.\,\ref{Discussion_rim}. 

\begin{table*}
    \caption[]{Polarimetric properties of the sample.}
    \label{Pol_properties}
    \centering
    \begin{tabular}{lcccccl}
        \hline
        \hline
        \noalign{\smallskip}
        Source & $r$ (\arcsec) & P.A.\ (\degree) & $\delta_{\rm pol}$ (10$^{-3}$) & DoLP (\%) & AoLP (\degree) & Remarks \\
        \noalign{\smallskip}
             \hline
        CI Tau & & & 1.3$\pm$0.3 & 0.56$\pm$0.06 & 77.9$\pm$2.8 \\
        HD287823 & & & 0.5$\pm$0.4 & 0.16$\pm$0.12 & 87.6$\pm$42.5 \\
        HD287823 B & 2.32 & 329 & - & <0.6 & - & Upper limit on DoLP is not very constraining. \\
        HD245185 & & & 1.0$\pm$0.2 & 1.14$\pm$0.13 & 169.6$\pm$3.5 \\
        HD290770 & & & <0.2 & 1.00$\pm$0.04 & 88.4$\pm$1.1 & Possible circumstellar unresolved polarization. \\
        TW Cha & & & 0.5$\pm$0.3 & 0.43$\pm$0.01 & 148.3$\pm$0.4 \\
        DI Cha & & & <0.3 & 1.42$\pm$0.03 & 143.7$\pm$1.1 & Large $A_{\rm V}$. DoLP may be due to interstellar material. \\
        DI Cha BC & 4.58 & 201 & - & 1.3$\pm$0.6 & 141$\pm$2 & Large $A_{\rm V}$. DoLP may be due to interstellar material. \\
        DI Cha D & 0.22 & 245 & - & 1.3$\pm$0.1 & 137$\pm$2 & Large $A_{\rm V}$. DoLP may be due to interstellar material. \\
        HD98922  & & & 0.2$\pm$0.1 & 0.42$\pm$0.13 & 17.6$\pm$4.2 \\
        HD98922 B & 4.61 & 1 & - & 0.4$\pm$0.1 & 157$\pm$9 & Foreground object (Gaia EDR3). \\
        GW Lup & & & 0.4$\pm$0.2 & 0.18$\pm$0.11 & 161.6$\pm$42.0 \\
        HD142666 & & & 1.4$\pm$0.7 & 1.11$\pm$0.23 & 68.8$\pm$9.7 \\
        HD144432 & & & 0.5$\pm$0.3 & 0.43$\pm$0.17 & 12.2$\pm$40.6 \\
        HD144432 B & 1.43 & 6 & - & >0.7 & 153$\pm$3 & The stellar flux is saturated. \\
        HD144432 C & 0.08 & 336 & - & >0.7 & 145$\pm$4 & The stellar flux is saturated. $r$ and P.A.\ from B.  \\
        V1003 Oph & & & 0.2$\pm$0.1 & 0.48$\pm$0.05 & 163.3$\pm$1.9 \\
        WSB82 & & & 0.4$\pm$0.2 &3.55$\pm$0.05&56.4$\pm$0.5 & Large $A_{\rm V}$. DoLP may be due to interstellar material.\\
        HD150193 & & & <0.3 & 2.36$\pm$0.81 & 58.4$\pm$8.4 & Large $A_{\rm V}$. DoLP may be due to interstellar material. \\
        HD150193 B & 1.12 & 228 & - & 2.0$\pm$0.3 & 58$\pm$3 & Large $A_{\rm V}$. DoLP may be due to interstellar material. \\
        AK Sco & & & 1.5$\pm$0.8 &1.25$\pm$0.05&135.3$\pm$3.0\\
        HD179218 & & & 2.3$\pm$0.6 & 0.98$\pm$0.17 & 101.8$\pm$6.0 \\
        HD179218 B & 2.35 & 139 & - & 0.8$\pm$0.1 & 70$\pm$1 & Possible circumstellar unresolved polarization. \\
        HD179218 C & 3.75 & 338 & - & <0.5 & - & \\
        \hline
    \end{tabular}
     \tablefoot{Columns: source name, separation and position angle from the primary star (for companions only), resolved polarized-to-stellar light contrast (see Sect.\,\ref{Disk_brightness}), degree and angle of linear unresolved polarization (see Sect.\,\ref{Unresolved_pol}), remarks. The P.A.\ of a disk inducing the unresolved polarization is expected to be perpendicular to the AoLP. DI Cha B and C are not resolved in our image.}
   \end{table*}

\begin{figure}
  \centering
 \includegraphics[width=8cm]{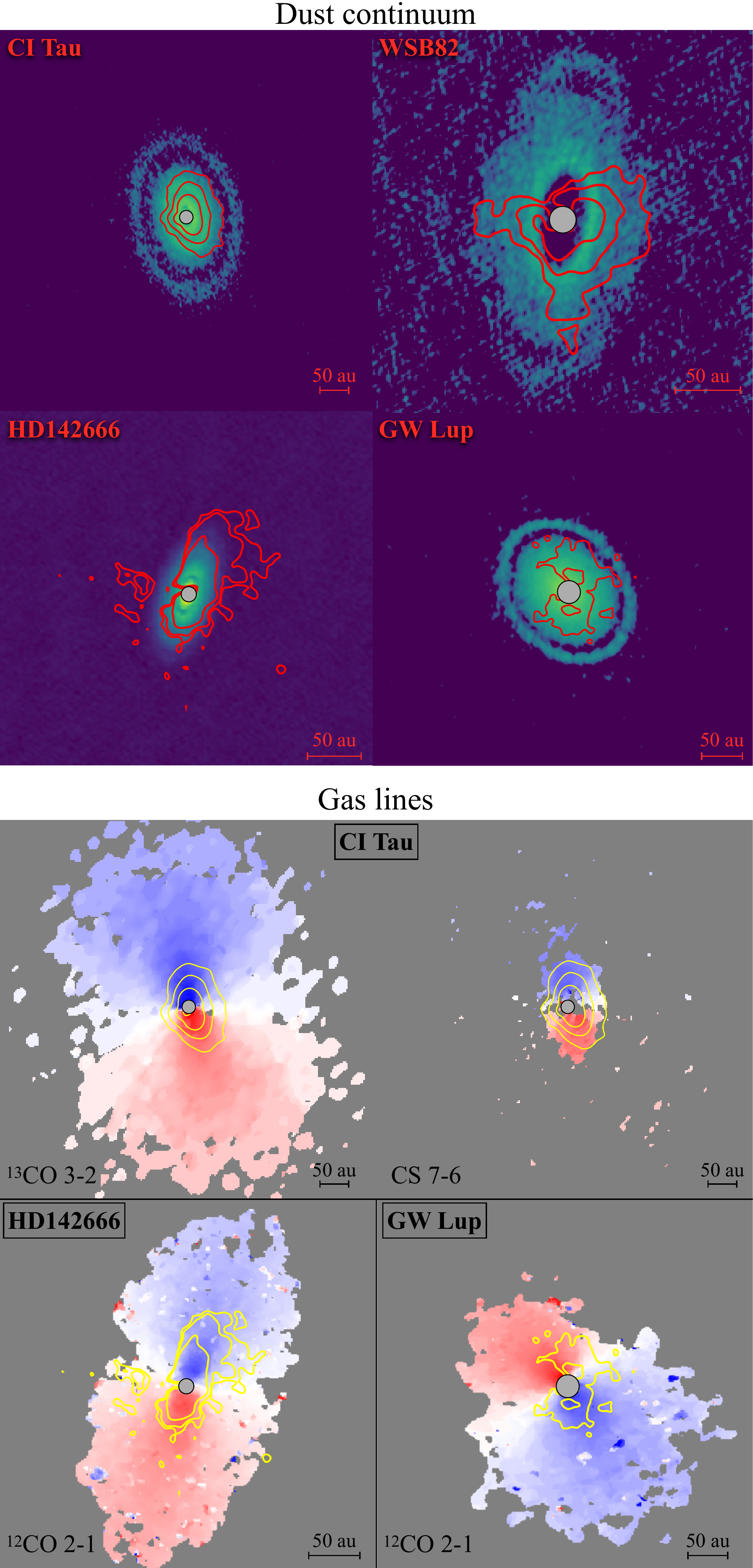} 
      \caption{Comparison with ALMA high-resolution images. The continuum maps at 1.3mm are shown and the top, and the intensity-weighted mean velocity (moment 1) maps clipped at $3\sigma$ of varied gas lines are at the bottom. {The $\sigma$ level is determined from a region as large as a resolution element devoid of signal in the channel maps.}  In each panel, the SPHERE detection at 3$\sigma$, 9$\sigma$, and 27$\sigma$ significance is shown with contours. The CI Tau continuum image is from \citet{Clarke2018}, while the $^{13}$CO and CS images are from \citet{Rosotti2021}. The continuum and the $^{12}$CO maps of GW Lup and HD142666 are from the DSHARP program \citep{Andrews2018}. The continuum image of WSB82 is from the ODISEA program \citep{Cieza2021}. Images from the same target are shown with the same spatial scale. The inner gray region indicates the SPHERE coronagraph. North is up and east is left.   
       }
          \label{ALMA_comparison}
  \end{figure}

\subsection{Stellar companions} \label{Stellar_companions}

Several stellar companions are visible from the SPHERE total intensity images. Here, we focus on those that are bright enough (1\% of the primary flux) to allow for a polarimetric characterization. None of these companions show resolved polarized signal but most of them show evidence of unresolved polarized light. To characterize this component, we followed the procedure highlighted by \citet{vanHolstein2021}. We computed the total flux in an 8-pixel-wide aperture around the companion in the $Q$ and $U$ images as well as in their respective intensity images $I_Q$ and $I_U$. To constrain the background flux, we performed the same computation on several positions at the same separation from the main star, excluding those affected by diffraction spikes or bad pixels. After subtracting the resulting background flux, we computed $q$, $u$, DoLP, and AoLP as described in Sect.\,\ref{Observations}. For HD150193, we used the non-coronagraphic dataset (see Appendix \ref{appendix_observations}) where the companion is not saturated. The resulting DoLP and AoLP for all bright stellar companions are listed in Table \ref{Pol_properties} along with their astrometrical properties.

HD287823 B is detected to the north of the primary at a projected distance of 800 au. The parallax of these two stars from Gaia EDR3 is the same \citep{Gaia2021}, suggesting their bound nature. No sign of polarization was found in HD287823 B, but the upper limit that we derived (0.6\%) is not particularly meaningful. 

The total intensity image of DI Cha reveals both BC and D \citep{Lafreniere2008} but the individual components B and C are not resolved in our image. From both BC and D, we inferred the same degree and angle of polarization of A, confirming that the polarization of all stars is induced by material in the line of sight (see Sect.\,\ref{Unresolved_pol}). 

According to Gaia EDR3, HD98922 B is a foreground object (300 vs. 650 pc). The degree of polarization is comparable to HD98922, but the angle is different. Furthermore, the AoLP of HD98922 is coherent with the resolved disk P.A.\ (see Sect.\,\ref{Unresolved_pol}) and the interstellar extinction is very low (see Table \ref{Stellar_properties}). Therefore, HD98922 B is likely to have its own polarization feature.

HD144432 B and C are clearly resolved in our images. Excluding the innermost saturated region, the southeastern star in the image is 15$-$20\% brighter than the companion and is therefore B, following \citet{Mueller2011}. Their saturated flux only allows us to infer an upper limit to their DoLP. Interestingly, this is larger than the DoLP inferred for the primary. Considering this discrepancy as well as the low interstellar extinction ($A_{\rm V}$=0.4 mag), we conclude that the observed polarization is produced locally and is thus intrinsic of each star. Nonetheless, the calculation of B and C may not be accurate as the two stars are very close to each other and their flux on the detector is saturated. HD150193 B \citep{Bouvier2001} has a comparable degree and equal angle of polarization to the primary, suggesting that the polarization is induced by interstellar material. 

Finally, HD179218 B seems to have its own polarization feature with a high DoLP (though comparable with the primary) and a different AoLP than the primary. This conclusion is supported by the DoLP of a third star in the field \citep[\mbox{HD179218 C},][]{Fukagawa2010}, exhibiting a DoLP lower than those of both A and B (<0.5\%).

\section{Comparison with ALMA images} \label{ALMA}
Ten of the fifteen targets of this work have been observed with ALMA {(see Appendix \ref{appendix_observations})}. Five of these have been imaged with high angular resolution (<0.1\arcsec), enabling direct comparison with the SPHERE images.

\subsection{High-resolution images} \label{ALMA_high}
The continuum and line images of the four targets with published work are compared with the SPHERE detection isophotes in Fig.\,\ref{ALMA_comparison}. The continuum Band 6 maps of CI Tau, WSB82, and GW Lup \citep[respectively]{Clarke2018, Cieza2021, Andrews2018}
show several disk substructures (rings in all three targets and an inner cavity in WSB82) that do not have any visible counterpart in the SPHERE images. They also reveal that the scattered-light emission profile {probed by observations carried out with standard integration times (see Appendix \ref{appendix_observations})} is compact compared to the thermal emission from millimeter grains. This is in principle surprising since the scattered light indirectly traces the gaseous disk extent, and it is typically detected at larger radii than the ALMA continuum emission \citep[see e.g.,][]{vanBoekel2017, Garufi2020a}. This incongruity is also evident when comparing the signal in scattered light with the molecular line emission. 

The $^{13}$CO line emission from CI Tau by \citet{Rosotti2021} is detected at radii approximately four times larger than the scattered light (see Fig.\,\ref{ALMA_comparison}). These authors also revealed the presence of a gap in $^{13}$CO at a separation of 50 au and concluded that it is, to some extent, a real effect caused by shadowing by the inner disk regions. This gap is not visible in scattered light and this is possibly due to the different vertical heights $h/r$ of {$^{13}$CO and scattered light}. In fact, the CS emission from the same authors does not exhibit any gap at 50 au either. Figure \ref{ALMA_comparison} also shows a comparison between the SPHERE isophotes and the CS emission, revealing that the two components are similarly distributed in both outer extent and vertical origin (showing the same offset along the disk minor axis). \citet{Rosotti2021} constrained a similar $h/r$ for $^{13}$CO and CS (approximately 0.1), but it is still plausible that the $h/r$ of CS is larger \citep[see e.g.,][]{LeGal2019, Podio2020} and therefore closer to the $h/r$ that is typically measured for the NIR scattered light \citep[0.15$-$0.25,][]{Ginski2016, Avenhaus2018}.

The peculiar distribution of the faint scattered light around GW Lup (one lobe to the NW and one lobe to the S) can also be investigated through the comparison with the $^{12}$CO line by the DSHARP project \citep{Andrews2018}. Interestingly, the emission from this line (which is mostly sensitive to the disk temperature) appears more extended to the {blueshifted} disk side, which is where the southern SPHERE lobe resides. The other scattered-light lobe may correspond to the forward peak of the scattering phase function since the NW disk side is most likely the near side (the $^{12}$CO emission in fact bends toward the SE). A similarly co-spatial distribution between the scattered light and the $^{12}$CO is also seen in WSB82 (Gonz\'{a}lez-Ruilova et al.\,in prep.). The western near-IR lobe may trace the disk's near side, while the southern lobe may correspond to a local increase of $^{12}$CO emission.  

The geometrical interpretation of the disk around HD142666 is complicated by the asymmetry shown in scattered light by the northern and southern disk sides. The near-IR signal to the north is more extended than the continuum emission, while the signal to the south is detected to a similar separation. Speculatively, this discrepancy could be due to a different illumination of the two disk sides, similarly to HD144432 and V1003 Oph (see Sect.\,\ref{Individual}), {or to an actual asymmetric distribution of dust material \citep{Dong2015b} In any case}, the $^{12}$CO emission from this source does not reveal any obvious asymmetry. 

Finally, AK Sco was imaged multiple times with increasing resolution. These images helped constrain the coplanarity of gas dynamics and binary system orbit to within a few degrees \citep{Alencar2003, Czekala2015, Czekala2019}. The latest high-resolution continuum image taken in Band 6 constrains the presence of a large disk cavity (approximately 20 au) and of faint unresolved emission centered on the binary (I.\,Czekala, private comm.) that could, speculatively, be related to the shadow seen in scattered light (see Sect.\,\ref{Individual}). Within the sample of this work, the presence of a large cavity is only inferred in this source and in WSB82, and it is discussed in Sect.\,\ref{Discussion_rim}.

\subsection{Moderate-resolution images}
HD245185, TW Cha, DI Cha, HD98922, and V1003 Oph have only been observed with moderate angular resolution (0.2\arcsec$-$0.8\arcsec). Nevertheless, these images offer some useful constraints to interpret the SPHERE images.

HD245185 was observed in Band 6 in the context of a survey of $\lambda$ Orionis cluster \citep{Ansdell2020}. The disk turned out to be by far the brightest of the {ALMA} sample. \citet{Ansdell2020} constrained the continuum emission extent by means of elliptical gaussian axes of 0.20\arcsec\ and 0.11\arcsec . The authors also present the $^{12}$CO map of the source, finding that the 3$\sigma$ detection significance of this line is met at a separation of 0.8\arcsec. Therefore, the detectable scattered-light outermost radius (0.5\arcsec\ corresponding to 200 au) is, unlike the cases shown in Sect.\,\ref{ALMA_high}, intermediate between the dust and gaseous disk extent.

TW Cha and DI Cha were observed in Band 7 in the context of a survey of the Chamaeleon-I star-forming region \citep{Pascucci2016}. These authors fit the visibility data, finding semi-axes of 0.40\arcsec $\times$0.33\arcsec\ and 0.29\arcsec $\times$0.29\arcsec\ that translate to disk sizes of 70 au and 55 au, respectively. Therefore, similarly to the targets of Fig.\,\ref{ALMA_comparison}, the {observed} scattered-light emission from both sources is less extended than the continuum emission.

A simple Gaussian fit using the \textit{uvmodelfit} task with the Common Astronomy Software Applications package \citep[CASA,][]{McMullin2007} was performed on the continuum emission from archival maps of HD98922 and V1003 Oph (PIs: Panic O.\, and Du F., respectively). The disk around the former turned out to be unresolved, yielding a loose constraint of less than 500 au in size. On the other hand, the emission around V1003 Oph is resolved (0.43\arcsec$\times$0.32\arcsec), indicating a disk extent of at least 50 au, which is once again larger than what is constrained with SPHERE (see Fig.\,\ref{Imagery}).

\begin{figure}
  \centering
 \includegraphics[width=9cm]{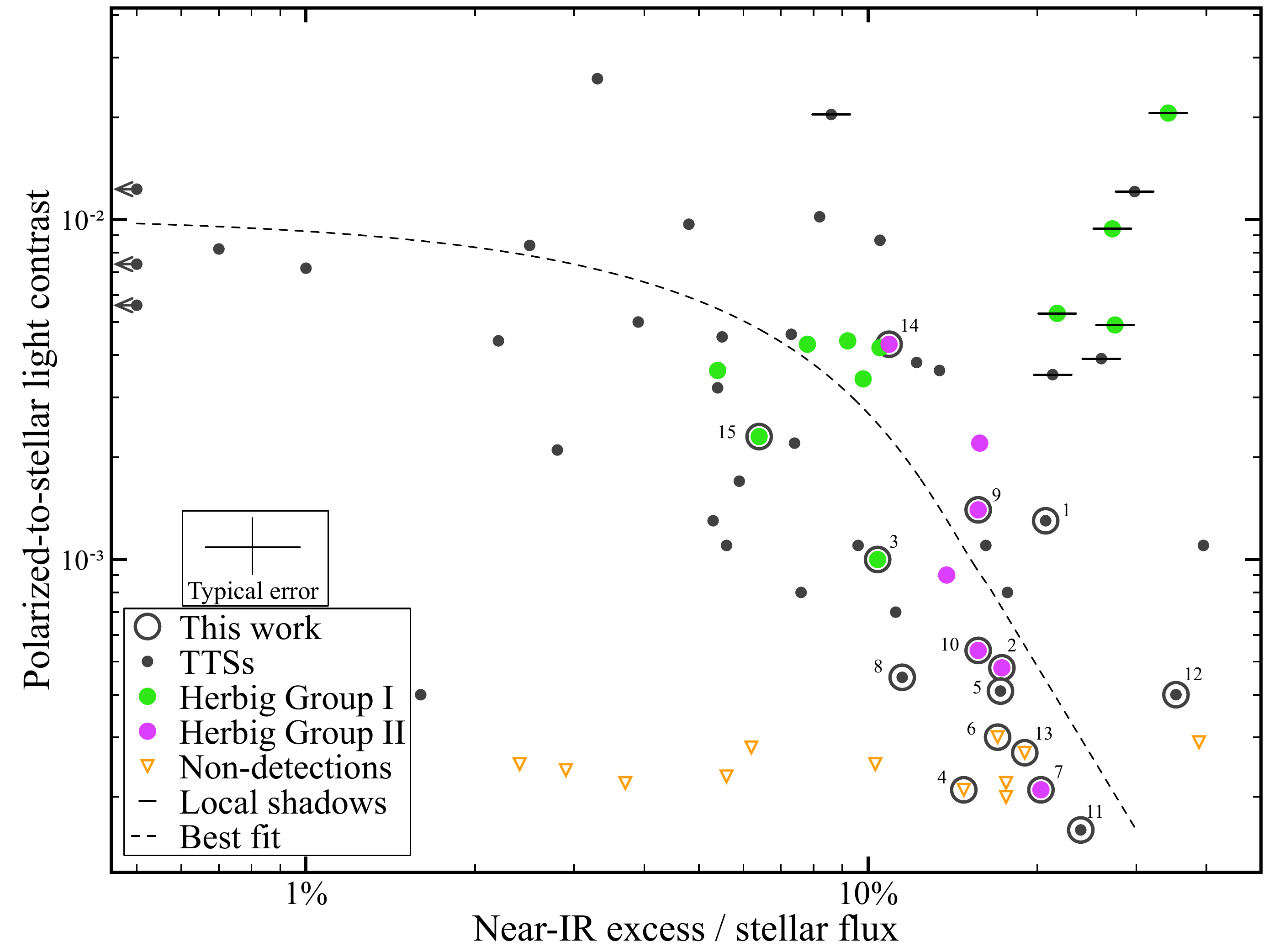} 
      \caption{Disk brightness in scattered light vs. near-IR excess. The polarized-to-stellar light contrast for the targets of this work and benchmark sample (see Sect.\,\ref{Disk_brightness}) is compared with the near-IR excess from the SED (see Appendix \ref{appendix_properties}). The spline fit is performed on all sources except non-detections and targets with local shadows (see Sect.\,\ref{correlation_NIR}). Target numbering refers to Table \ref{Stellar_properties}.}
          \label{Contrast_NIR}
  \end{figure}

\section{Search for trends} \label{Correlations}
In this section, we investigate the peculiar faintness in scattered light of the main sample of Fig.\,\ref{Imagery} by relating the polarized light contrast {of the objects shown} in Fig.\,\ref{Contrast_list} with other disk and stellar properties. Given the bias on very inclined disks (see Sect.\,\ref{Disk_brightness}), J1608-382, RY Lup, MY Lup, T Cha, AK Sco, and HD142666 are excluded from the analysis.

\subsection{NIR excess} \label{correlation_NIR}
The interplay between the inner and outer disk is investigated in Fig.\,\ref{Contrast_NIR} by comparing the integrated NIR excess from the SED (see Appendix \ref{appendix_properties}) with the disk brightness in scattered light {expressed through the contrast $\delta_{\rm pol}$ of Sect.\,\ref{Disk_brightness}}. In the diagram, we label those disks that show shadows and spirals from the scattered-light images. It is well established that sources with spiral disks all exhibit an anomalously high NIR excess as well as some other peculiar properties: a larger metallicity than the other GI \citep{Kama2015}, a lower CO vibrational ratio \citep{Banzatti2018}, localized shadows when a NIR cavity is present \citep{Garufi2018}, and azimuthal asymmetries from millimeter continuum images \citep{vanderMarel2021}. In view of the extraordinary origin of this NIR flux, we exclude these sources {as well as the non-detections in scattered light} from the search for trends with the disk brightness. By doing this, the two quantities show a clear, though mild, anticorrelation. The Pearson correlation coefficient is $-$0.49, and a {cubic} spline interpolation yields the fit shown in Fig.\,\ref{Contrast_NIR}. The trend is particularly evident when focusing on the targets of this work and suggests that the disk faintness in scattered light of the sources in question is intimately related to the geometry of the inner disk, {and this is further discussed in Sect.\,\ref{Discussion_rim}}.

Another interesting peculiarity of Fig.\,\ref{Contrast_NIR} is the presence of six non-detections from the literature {\citep{Garufi2020a}} with NIR excess below 10\%. These sources clearly fall far from the interpolation fit. The most likely explanation for this incongruity is that these sources are small. This view is supported by their low disk mass (see Sect.\,\ref{correlation_mass}).

\begin{figure}
  \centering
 \includegraphics[width=9cm]{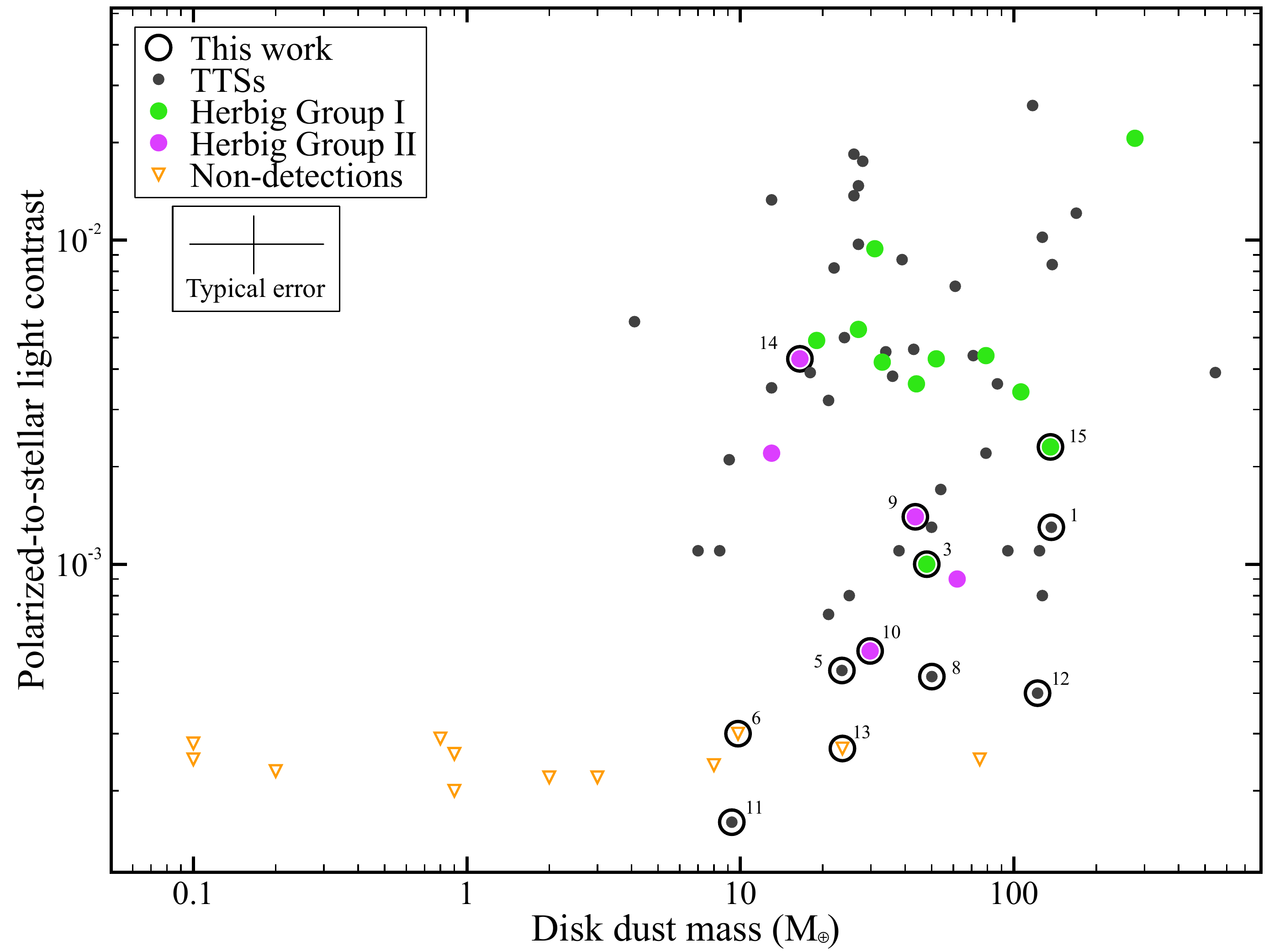} 
      \caption{Disk brightness in scattered light vs. disk mass. The polarized-to-stellar light contrast for the targets of this work and benchmark sample (see Sect.\,\ref{Disk_brightness}) is compared with the dust content in disks calculated from the millimeter integrated flux (see Appendix \ref{appendix_properties}). Target numbering refers to Table \ref{Stellar_properties}. {Missing targets have no mass estimate available.}}
          \label{Contrast_mass}
  \end{figure}

\subsection{Disk dust mass} \label{correlation_mass}
The relation between the {contrast $\delta_{\rm pol}$} and the disk dust mass {calculated as described in Appendix \ref{appendix_properties}} is investigated in Fig.\,\ref{Contrast_mass}. What clearly stands out from the diagram is that the majority of the undetected disks from the literature (8 out of 12, {see \citealt{Garufi2018, Garufi2020a}}) have significantly lower mass than the detected disks (<10 M$_\oplus$). These eight targets include those that fall out of the correlation found with the NIR flux in Fig.\,\ref{Contrast_NIR}, corroborating the thesis that these disks are small. {Instead}, the two non-detections from this work with millimeter flux available (DI Cha and HD150193, numbers 6 and 13) sit in the lower end of the mass distribution of detected disks implying that their disk could still be self-shadowed rather than small (as also suggested by their position in Fig.\,\ref{Contrast_NIR}).

As for the detections, the diagram of Fig.\,\ref{Contrast_mass} does not reveal any correlation between disk brightness in scattered light and disk mass. In particular, the faint disks of this work exhibit a median disk mass in line with the benchmark sample {of Fig.\,\ref{Contrast_list}} (55 vs 65 M$_\oplus$). This indicates that the illumination of the disk surface has no relation with the solid content in the disk.

\subsection{Stellar age} \label{correlation_age}
\begin{figure}
  \centering
 \includegraphics[width=9cm]{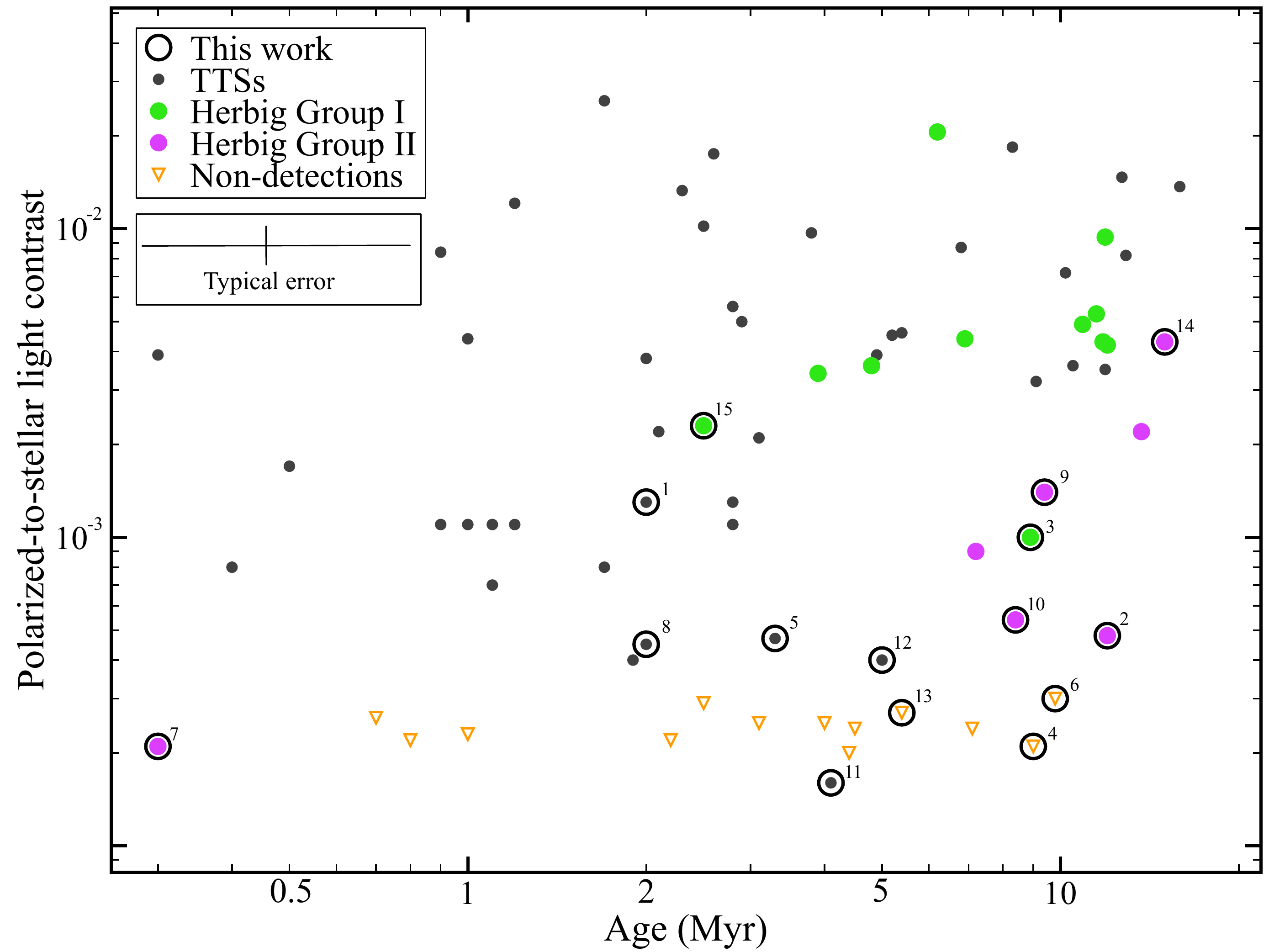} 
      \caption{Disk brightness in scattered light vs. age. The polarized-to-stellar light contrast for the targets of this work and benchmark sample (see Sect.\,\ref{Disk_brightness}) is compared with the stellar age constrained from evolutionary pre-main-sequence tracks (see Appendix \ref{appendix_properties}). Target numbering refers to Table \ref{Stellar_properties}.}
          \label{Contrast_age}
  \end{figure}

The temporal evolution of the disk brightness in scattered light is evaluated in Fig.\,\ref{Contrast_age}. Even though no obvious trend is visible from the diagram, young sources ($<$2 Myr) are found to be on average fainter than old sources (3.5 $\times$ 10$^{-2}$ and 5.6 $\times$ 10$^{-2}$, respectively). This behavior is consistent with what was found by \citet{Garufi2018}. The future release of observations of TTSs from a less biased sample from young star-forming regions \citep[as in the ongoing program DESTINYS,][]{Ginski2020} may accentuate the observed trend for old disks to be brighter in scattered light.

\section{Discussion} \label{Discussion}
\begin{figure*}
  \centering
 \includegraphics[width=18cm]{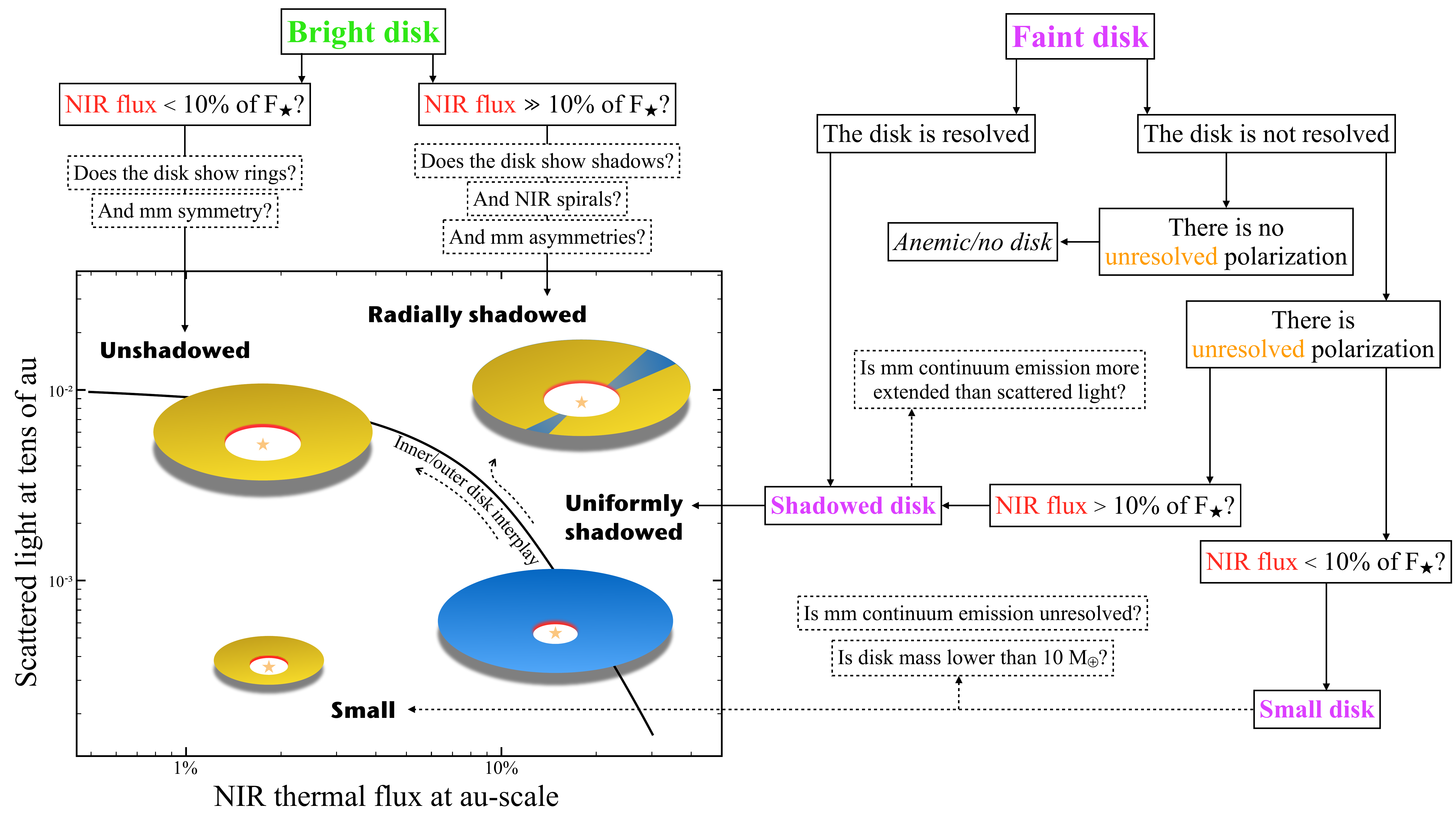} 
      \caption{Workflow for the interpretation of scattered-light imaging based on the results of this work on faint disks and those of \citet{Garufi2018}, \citet{Banzatti2018}, and \citet{vanderMarel2021} concerning bright disks. {The sketch in the diagram refers to the result of Fig.\,\ref{Contrast_NIR}.}}
          \label{Workflow}
  \end{figure*}

Planet-forming disks that are faint or undetected in scattered light represent a poorly explored sample that is, however, pivotal when it comes analyzing disk demographics. The detection of a disk in scattered light can be hindered by a limited {disk} size, poor illumination, {as well as a small albedo}. In Fig.\,\ref{Workflow}, we provide a workflow to interpret the observations. The observed anticorrelation between the disk brightness in scattered light and the near-IR excess from the SED (Fig.\,\ref{Contrast_NIR}) suggests the following interpretation. 
 
First, a NIR excess of 10\%$-$20\% of the stellar flux is indicative of a full inner rim that casts a uniform shadow on the outer disk. These objects (such as HK Lup and GW Lup) are the self-shadowed disks predicted by \citet{Dullemond2001}.
 
Second, smaller values (<10\%) suggest a partly depleted inner disk, as corroborated by the common detection of a resolved large cavity. These objects (such as PDS70 and HD169142) are commonly called transition disks and show no sign of shadows from the outer disk.
 
Third, larger values (>20\%) suggest an additional source of reprocessed starlight by the disk's inner rim, despite the presence of a resolved large cavity. These objects (such as HD142527 and HD135344B) are the transition disks that show radial shadow lanes from the outer disk {cast by a misaligned inner disk}.

The finite spatial resolution and the use of a coronagraph determine an additional cause of non-detections, such that small values of NIR excess (<10\%) associated with a non-detection in scattered light are indicative of a small disk (<20 au) with a partly depleted inner region. These objects (such as DoAr16 and SR9) may be the compact counterpart of the large transition disks, but they are still poorly known.

All the targets of this work {follow the anti-correlation between NIR excess and polarized contrast. Furthermore, a limited disk size as the main explanation of their faintness in scattered light is ruled out by their disk dust mass being similar to that of bright disks (see Sect.\,\ref{correlation_mass}). These two elements suggest that they} belong to the category of self-shadowed disks. {This is not surprising since} the whole sample was selected based on the SED, and a star hosting a self-shadowed disk always exhibits the SED of Herbig Ae/Be (or TTS) Group II. The inner rim of these disks intercepts a lot of starlight and prevents the direct illumination of the outer disk. In turn, the poorly illuminated outer disk limits both the emission of FIR thermal light and the scattering of NIR light. These two quantities are in fact known to {positively} correlate \citep{Garufi2017}.

\subsection{Inner rim and outer disk interplay} \label{Discussion_rim}
The location of the disk inner rim is determined by several dust, disk, and stellar properties \citep[see e.g.,][]{Isella2005, Kama2009}. NIR interferometric observations of Herbig stars show that the inner rims are smooth, radially extended, and sometimes azimuthally asymmetric structures \citep{Lazareff2017, Perraut2019, Kluska2020}. Coarsely speaking, the interval of stellar luminosities in our benchmark sample translates to inner rims expected to be located from 0.02 au to 2 au \citep[see e.g.,][]{Millan-Gabet2007}. Nevertheless, we found no clear dependency for the outer disk brightness on the spectral type (see Fig.\,\ref{Contrast_list}). This means that the interplay between the location of the inner rim, the extent and depth of the shadow cast, and the outer disk flaring remain largely unaltered in the available sample of M4$-$B9 spectral types. 

The relative alignment of the inner rim and outer disk is also fundamental to determining the type and amount of shadow cast. Numerical simulations showed that even minor misalignments (of the order of the degree) can induce an azimuthally confined, though broad, shadow on the outer disk \citep[see e.g.,][]{Nealon2019, MuroArena2020}. Therefore, all disks with uniformly faint signal (see e.g., HD245185 and TW Cha from Fig.\,\ref{Imagery}) are not expected to host any misaligned inner disk. On the other hand, disks with clear azimuthal variations such as V1003 Oph and WSB82 may do so. At present, such small misalignments cannot be directly measured as the uncertainties involved in the near-IR interferometric imaging \citep{Perraut2019} and in the characterization of the unresolved polarization (see Sect.\,\ref{Unresolved_pol}) are large. 

While the scattering surface of bright disks is well studied, that of faint disks is barely accessible with the sensitivity of the current sample. Vertical heights $h/r$ of 0.15$-$0.25 have been constrained from bright disks \citep{Ginski2016, Avenhaus2018}. Smaller heights are expected from shadowed disks \citep{Dullemond2001}, but their direct measurement is entrusted to future, considerably deeper observations of this type of disk. Furthermore, none of our images show an abrupt change in the radial brightness distribution. In principle, this may suggest that the disk never emerges from the shadow so as to continue as a flaring disk outward, as suggested by \citet{Dullemond2001}. It is, however, possible that this transition occurs very smoothly because of the gentle decrease of optical depth with increasing height expected for a realistic inner disk rim, as demonstrated by \citet{Dong2015c}.

As of today, the vertical stratification of different molecular species and their relation with the scattered light are poorly known. {However,} this may be pertinent to disk shadowing. {In fact,} the emission of optically thick molecules sensitive to the disk temperature might show gradients or even sub-structures that are not determined by the local gas column but rather by some shadowing effect \citep{Rosotti2021}. The spatial analogy of the scattered light and molecular emission (in particular the CS in CI Tau) shown in Fig.\,\ref{ALMA_comparison} suggests an intimate relation between the gaseous emission and the illumination pattern. Thus, NIR imaging of the scattered light offers a fundamental support to the study of disk line emission that has not been fully exploited yet.  

\subsection{Faint disk demographics and evolution}\label{Discussion_evolution}

Both faint and bright disks that are smaller than $\sim$20 au cannot be resolved by NIR imaging. The incidence of such small disks assessed from ALMA surveys is high \citep[see e.g.,][]{Ansdell2016, Barenfeld2016, Cieza2019}. At present, disks less massive than 10 M$_\oplus$ in dust are not resolved \citep[see Fig.\,\ref{Contrast_mass} and][]{Garufi2020a}. Nevertheless, their detection and characterization in the NIR is becoming feasible through their induction of unresolved polarization \citep[see Sects.\,\ref{Unresolved_pol}, \ref{Stellar_companions}, and][]{vanHolstein2021}. 

Contrary to small disks, the incidence of self-shadowed disks on the disk demographics can only be evaluated from a less biased near-IR census of the nearby star-forming regions. Meanwhile, the clear association between Group II targets and self-shadowed disks, as well as the extension of the \citet{Meeus2001} classification to TTSs, allows us to evaluate the frequency of shadowed disks on a much larger sample and to study the intermediate cases between shadowed and bright disks. In fact, TTSs are more common and they probe younger stellar ages as well as larger stellar mass intervals than Herbig AeBe stars. In the NIR/FIR diagram of Fig.\,\ref{FIR_NIR}, several TTSs lie between the parameter spaces of Group II and Group I. Targets between the Group II regime ({quadrant Q4}) and the low-NIR Group I ({Q2}) may host a shallow cavity that is potentially deepening with time. In the sketch of Fig.\,\ref{Workflow}, such targets (HD245185, GW Lup, and HD179218 from this work) would turn from a uniformly shadowed disk to an unshadowed disk moving upward along the interpolation line. In principle, AK Sco also belongs to this transition region. However, the presence of a large cavity from ALMA (see Sect.\,\ref{ALMA_high}) would suggest that this disk is a fully-fledged Group I, and that the low FIR excess and moderate amount of scattered light (see Sect.\,\ref{Disk_brightness}) is related to the very old age of the star (>12 Myr) pointing to a late evolutionary stage of the disk.  

The interpretation of the transition between Group II and high-NIR Group I ({Q1} in Fig.\,\ref{FIR_NIR}) is intimately related to the origin of the very high NIR excess exhibited by the latter sources \citep[see][]{Garufi2017, Banzatti2018}. The striking analogy with the presence of localized shadows in the outer disk suggests that this excess originates from inner material in a peculiar morphology that can cast radial shadow lanes rather than uniform shadows. This material could be an inclined broken disk or warp \citep[see e.g.,][]{Marino2015, Nealon2019, MuroArena2020}, a vortex of increased disk thickness at the inner edge of a dead zone \citep{Flock2017}, or a dusty wind in proximity to the dust sublimation radius \citep{Bans2012}. Intriguingly, two of the targets lying in the aforementioned transition space of Fig.\,\ref{FIR_NIR} (V1003 Oph and WSB82) show local shadows in our images. These sources may eventually leave the interpolation line of Fig.\,\ref{Workflow} to turn into bright, radially shadowed disks, such as the prototypical examples of HD100453 and HD143006 \citep{Benisty2017, Benisty2018}. The presence of a disk cavity imaged in WSB82 \citep[see Fig.\,\ref{ALMA_comparison} and][]{Cieza2021} also supports the intermediate nature of this target. Finally, another source sitting between Group II and high-NIR Group I (HD287823) shows the possible presence of a small inner cavity from our images (see Sect.\,\ref{Unresolved_pol}) and thus represents an additional, possible link between the two groups.

\section{Summary and conclusions} \label{Conclusions}
Much of the focus of the planet formation community has thus far been put on extended and bright planet-forming disks with large inner cavities. Smaller and fainter disks are nonetheless fundamental to the disk demography. Disks that are faint in scattered light are known to be associated with Herbig Ae/Be Group II targets \citep{Grady2005, Garufi2017}, namely sources with low far-IR excess from the SED \citep{Meeus2001} and no large disk cavities \citep{Maaskant2013}. In this work, we extend the concept of Group I and Group II to TTSs and present the unpublished scattered-light images from VLT/SPHERE of 15 Herbig Ae/Be and TTSs Group II or I/II intermediate targets. 

All the disks into question turned out to be 1$-$2 orders of magnitude fainter {in scattered light} than a benchmark sample of known disks, in agreement with previous work on Group II sources. For the sources with ALMA high-resolution images available, the outer radius of the detectable scattered light is even smaller than the dust continuum, suggesting that the SPHERE images only probe a small portion of the disk extent. {Clearly, deeper observations of the scattered light may be able to retrieve signal from larger disk extents but the need for integration times longer than those that are typically adopted is a direct consequence of the disk faintness.}

The disk brightness calculated on the large benchmark sample with SPHERE observations shows a clear though loose anti-correlation with the near-IR excess from the SED. This suggests that the peculiar faintness in scattered light of Group II sources is due to the self-shadowed nature of these disks. Much of the starlight is reprocessed by the disk's inner rim at (sub-)astronomical unit scales, and the outer disk  is in penumbra at tens of astronomical units. We also report a possible spatial relation between the emission of some molecular lines and the scattered light, suggesting a dependency for the detectable line flux on the global illumination pattern of the disk.

We found no trend for the disk brightness of the large sample with the spectral type within the M4$-$B9 range. This implies that the amount of self-shadowing does not strictly relate to the location of the disk inner rim. There is also no clear relation between the disk brightness and the dust mass. Disks less massive than 10 M$_\oplus$ are undetected in scattered light but this is explained by their limited size rather than a shadowing effect. {However, all the disks presented in this work are massive enough to be detected in scattered light, reinforcing the idea that they are self-shadowed.} 

Owing to the observed weakness of temporal evolution for the median disk brightness, the evolutionary link of the self-shadowed disks with the brighter, well-studied disks with cavities remains speculative. The presence of substructures inferred from the ALMA continuum images of the self-shadowed disks may suggest that planet formation occurs in a similar manner. Whether all or some of these extended self-shadowed disks will eventually evolve into cavity-hosting disks remains unknown. However, we propose that some of the faint disks studied here (such as HD245185, HD287823, and WSB82) are at an intermediate stage toward either the bright unshadowed disks (such as PDS70 and HD169142) or the bright disks that show azimuthally confined shadows on their surface (such as HD142527 and HD143006), representing a possible fork in the disk evolution that is determined by the geometry of the inner disk region.

\begin{acknowledgements}
We thank the referee for the constructive report. We thank G.\,Rosotti, C.\,Clarke, L.\,Cieza and the DSHARP team for sharing their ALMA data as well as I.\,Czekala for useful information on their unpublished ALMA dataset. SPHERE is an instrument designed and built by a consortium consisting of IPAG (Grenoble, France), MPIA (Heidelberg, Germany), LAM (Marseille, France), LESIA (Paris, France), Laboratoire Lagrange (Nice, France), INAF‚ Osservatorio di Padova (Italy), Observatoire de Gen\`{e}ve (Switzerland), ETH Zurich (Switzerland), NOVA (Netherlands), ONERA (France) and ASTRON (Netherlands) in collaboration with ESO. SPHERE was funded by ESO, with additional contributions from CNRS (France), MPIA (Germany), INAF (Italy), FINES (Switzerland) and NOVA (Netherlands).  SPHERE also received funding from the European Commission Sixth and Seventh Framework Programmes as part of the Optical Infrared Coordination Network for Astronomy (OPTICON) under grant number RII3-Ct-2004-001566 for FP6 (2004‚2008), grant number 226604 for FP7 (2009‚2012) and grant number 312430 for FP7 (2013‚2016). We also acknowledge financial support from the Programme National de Plan\`{e}tologie (PNP) and the Programme National de Physique Stellaire (PNPS) of CNRS-INSU in France. This paper also makes use of the following ALMA data: ADS/JAO.ALMA\#2015.1.01600.S and 2015.1.00847.S. ALMA is a partnership of ESO (representing its member states), NSF (USA) and NINS (Japan), together with NRC (Canada), MOST and ASIAA (Taiwan), and KASI (Republic of Korea), in cooperation with the Republic of Chile. The Joint ALMA Observatory is operated by ESO, AUI/NRAO and NAOJ. This research has made use of the VizieR catalogue access tool, CDS, Strasbourg, France (DOI: \url{10.26093/cds/vizier}). The original description of the VizieR service was published in A\&AS 143, 23. This work was supported by the PRIN-INAF 2016 "The Cradle of Life - GENESIS-SKA (General Conditions in Early Planetary Systems for the rise of life with SKA)", the project PRIN-INAF-MAIN-STREAM 2017 "Protoplanetary disks seen through the eyes of new-generation instruments", the PRIN-INAF 2019 "Planetary systems at young ages (PLATEA)",  the program PRIN-MIUR 2015 STARS in the CAOS - Simulation Tools for Astrochemical Reactivity and Spectroscopy in the Cyberinfrastructure for Astrochemical Organic Species (2015F59J3R, MIUR Ministero dell'Istruzione, dell'Universit\`{a}, della Ricerca e della Scuola Normale Superiore), the European MARIE SKLODOWSKA-CURIE ACTIONS under the European Union's Horizon 2020 research and innovation program through the Project "Astro-Chemistry Origins" (ACO), Grant No 811312, INAF/Frontiera (Fostering high ResolutiON Technology and Innovation for Exoplanets and Research in Astrophysics) through the "Progetti Premiali" funding scheme of the Italian Ministry of Education, University, and Research, as well as NSF grants AST- 1514670 and NASA NNX16AB48G, the Italian Ministero dell'Istruzione, Universit\`{a} e Ricerca through the grant Progetti Premiali 2012 - iALMA (CUP C52I13000140001). CD and CG acknowledge support from the NWO TOP grant "Herbig Ae/Be stars, Rosetta stones for understanding the formation of planetary systems", project number 614.001.552. T.H.\ acknowledges support from the European Research Council under the Horizon 2020 Framework Program via the ERC Advanced Grant Origins 83 24 28. 
\end{acknowledgements}

\bibliographystyle{aa} % style aa.bst
\bibliography{Reference} 
%\bibliography{27940}

\begin{appendix}

\section{Stellar and disk properties} \label{appendix_properties}
The stellar properties of the targets of this work (Table \ref{Stellar_properties}) as well as those of the benchmark sample are calculated as in \citet{Garufi2018} considering the distance $d$ from the newly available Gaia EDR3 \citep{Gaia2021}. We collected the complete SED of each source from the VizieR catalogue (DOI: \url{10.26093/cds/vizier}). We then scaled a PHOENIX model of the stellar photosphere \citep{Hauschildt1999} with the effective temperature $T_{\rm eff}$ available from the literature to both $d$ and the $V$ magnitude de-reddened by the extinction $A_{\rm V}$ calculated from the $V$, $R$, and $I$ wavebands. From the model, we calculated the stellar luminosity $L_*$. Uncertainties for the stellar luminosity are propagated from $d$, $A_{\rm V}$ (20\%), and $T_{\rm eff}$ ($\pm$100 K). The source is then put in the HR diagram and compared with the \citet{Siess2000}, \citet{Bressan2012}, \citet{Baraffe2015}, and \citet{Choi2016} stellar evolutionary tracks. The notion that the evolutionary models of nonmagnetic stars tend to underestimate the stellar masses below 1.2 M$_\odot$ \citep{Hillenbrand2004, Braun2021} motivated the use of the magnetic tracks by \citet{Feiden2016} for sources below this threshold. The mass and age interval shown in Table \ref{Stellar_properties} accounts for the diverse results yielded by different models. 

The NIR and FIR excesses are calculated by integrating the observed flux above the stellar model in the 1.2$-$4.5 and 22$-$450 $\mu$m wavelength intervals, respectively. {The error bars shown in Fig.\,\ref{FIR_NIR} are derived from the uncertainties on the stellar luminosity since the excesses are in fractions of $L_*$.} The dust masses in the disk are constrained from the 1.3 mm flux under standard assumptions \citep[optically thin emission, dust temperature of 20 K, and dust opacity by][]{Beckwith1990}. Given the complexity of the assumptions involved \citep[see e.g.,][]{Birnstiel2018, Zhu2019, Ribas2020}, these values are only meant as crude estimations to put the source into question. {Their errors are derived from the photometric and distance uncertainties.}

\section{Observing setup} \label{appendix_observations}
A summary of the observing program and setup for all the sources studied in this work is given in Table \ref{Table_observations}.

\begin{table*}
      \caption[]{Summary of observations.}
         \label{Table_observations}
         \centering
         \begin{tabular}{lccccccc}
            \hline
            \hline
            \noalign{\smallskip}
            Source & SPHERE ID (P.I.) & Waveband & Coronagraph & DIT (s) & $t_{\rm exp}$ (min) & Seeing (\arcsec) & ALMA ID \\
             \noalign{\smallskip} 
             \hline
             CI Tau & 0100.C-0452 (Benisty) & BB H & N\_ALC\_YJH\_S & 64 & 34.1 & 0.57 & 2016.1.01370 \\
     \hline 
      HD287823 & 0102.C-0165 (de Boer) & NB H  & None & 4 & 15.3 & 1.27 & - \\
     \hline
    HD245185 & 0102.C-0656 (Facchini) & BB H & N\_ALC\_YJH\_S & 64 & 34.1 & 0.46 & 2017.1.00466 \\
     \hline
     HD290770 & 0102.C-0165 (de Boer) & NB H & None & 2 & 17.3 & 0.77 & -  \\
     \hline
    TW Cha & 1104.C-0415 (Ginski) & BB H & N\_ALC\_YJH\_S & 64 & 59.7 & 0.58 & 2013.1.00437  \\
     \hline
    DI Cha & 198.C-0209 (Beuzit) & BB H & N\_ALC\_YJH\_S & 32 & 25.6 & 0.55 & 2013.1.00437  \\
     \hline
    HD98922 & 0102.C-0165 (de Boer) & BB H & N\_ALC\_YJH\_S & 4 & 13.3 & 0.85 & 2015.1.01600.S \\
     \hline
    GW Lup & 1100.C-0481 (Beuzit) & BB H & N\_ALC\_YJH\_S & 64 & 21.3  & 0.53 & 2016.1.00484.L \\
     \hline
    HD142666 & 096.C-0248 (Beuzit) & BB J & N\_ALC\_YJ\_S & 64 & 85.3 & 1.92 & 2016.1.00484.L \\
     \hline
    HD144432 & 097.C-0523 (Beuzit) & BB J & N\_ALC\_YJH\_S & 32 & 76.8 & 0.52 & - \\
     \hline
    V1003 Oph & 1100.C-0481 (Beuzit) & BB H & N\_ALC\_YJH\_S & 32 & 34.1 & 0.82 & 2015.1.00487.S  \\
     \hline
    WSB82 & 1100.C-0481 (Beuzit) & BB H  & N\_ALC\_YJH\_S & 64 & 21.3 & 0.48 & 2018.1.00028 \\
    \hline
    \multirow{3}{*}{HD150193} & 097.C-0523 (Beuzit) & BB J & N\_ALC\_YJH\_S & 16 & 70.4 & 0.43 & \multirow{3}{*}{-}\\
      & 0101.C-0464 (Benisty) & NB H & None & 12 & 24.0 & 1.05 \\
      & 0101.C-0383 (de Boer) & NB H & None & 0.837 & 16.0 & 1.10 \\
      \hline
    AK Sco & 097.C-0523 (Beuzit) & BB H & N\_ALC\_YJH\_S & 64 & 42.7 & 0.47 & 2018.1.01782 \\
     \hline
     HD179218 & 097.C-0523 (Beuzit) & BB J & N\_ALC\_YJH\_S & 32 & 42.7 & 0.80 & - \\
     \hline
     
     \end{tabular}
     \tablefoot{Columns: source name, program ID (principal investigator), waveband, coronagraph, detector integration time (DIT), total exposure time, average DIMM seeing.}
   \end{table*}

\section{Q and U images} \label{Appendix_QU}
\begin{figure*}
  \centering
 \includegraphics[width=16cm]{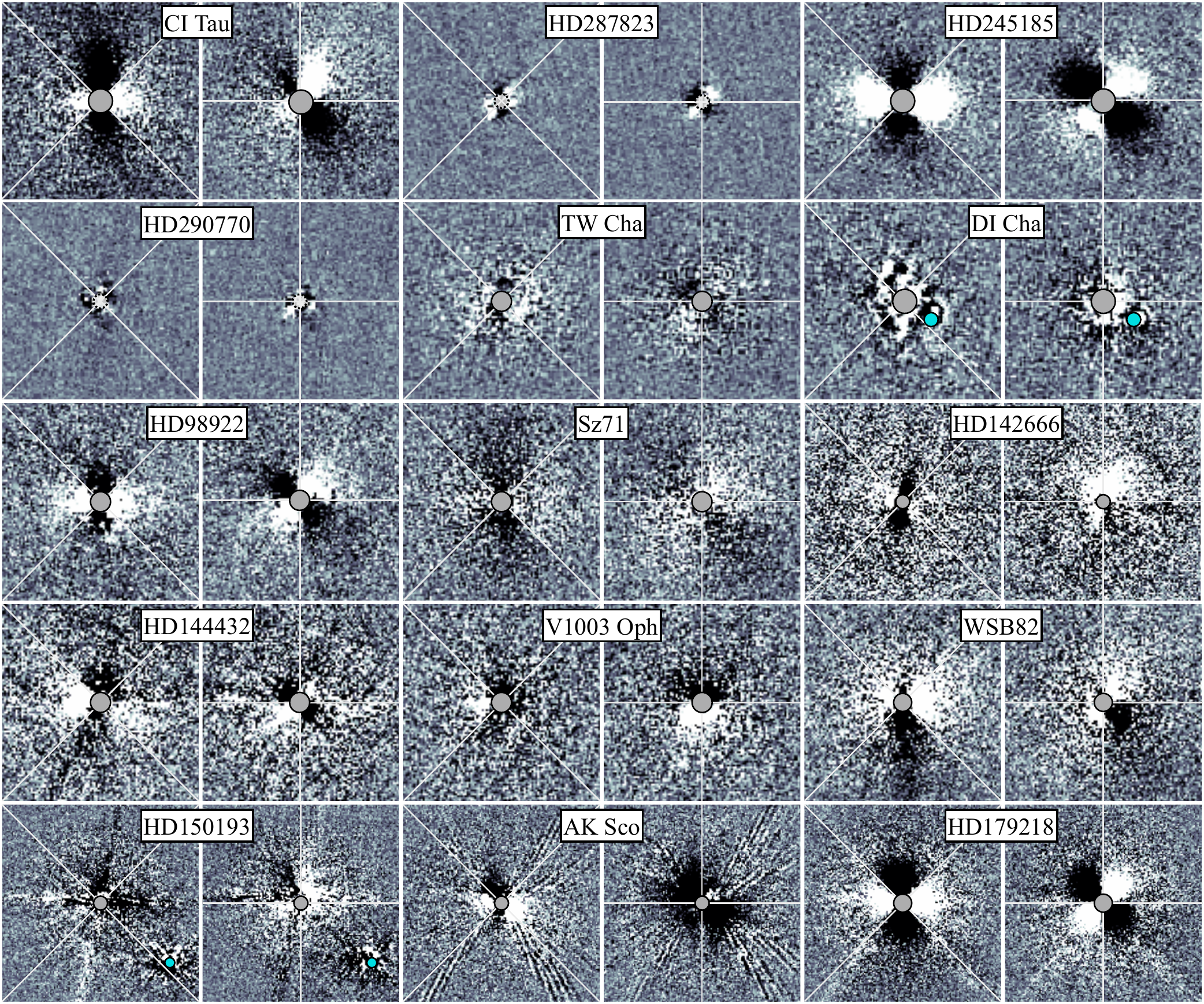}
      \caption{$Q$ and $U$ images of the whole sample. An ideal centro-symmetric scattering yields positive (or negative) signal to the east and west (or north and south) in the $Q$ images (left panels for each source), and to the SE and NW (or NE and SW) in the $U$ images (right panels), as is clear from e.g., HD245185 and HD179218.}
          \label{QU}
\end{figure*}

A simple practice to conclude whether weak signal from the $Q_\phi$ map is real polarized light is to search for a quadrupolar pattern in both the $Q$ and $U$ maps. In the ideal case of centro-symmetric single scattering, positive (or negative) signal is expected to the east and west (or north and south) in the $Q$ images, and to the SE and NW (or NE and SW) in the $U$ images. The $Q$ and $U$ images for the whole sample are shown in Fig.\,\ref{QU}. 

Some of the brightest disks in the sample (HD179218, HD245185, CI Tau, and HD98922) are a showcase of the practice. Some potentially dubious cases of Sect.\,\ref{Individual} (HD287823, TW Cha, GW Lup, WSB82) are well addressed by these images. The same is true for HD144432 and V1003 Oph if we focus on the eastern section only, which is where the putative polarized signal is detected in the $Q_\phi$ image, as well as for HD150193 if we focus on the SE-NW bisector. On the other hand, AK Sco and HD142666 exhibit a most peculiar pattern that is typical of inclined disks \citep[see e.g.,][]{Pohl2017}. Finally, HD290770 and DI Cha do not show any evidence of a centro-symmetric pattern. 

\end{appendix}

\end{document}